%% file: main.tex
\PassOptionsToPackage{dvipsnames,svgnames,table}{xcolor}

\documentclass[sigconf,nonacm,10pt]{acmart}

\setcopyright{none}
\acmConference[MobiCom '27]{The 33rd Annual International Conference on Mobile
Computing and Networking}{2027}{}
\acmBooktitle{}
\acmDOI{}
\acmISBN{}
\renewcommand\footnotetextcopyrightpermission[1]{}

\usepackage{bm}
\usepackage{enumitem}

\usepackage{subcaption} %
\usepackage[font=small]{caption}

\usepackage{booktabs}
\usepackage{calc}
\usepackage{array}
\usepackage{multirow}
\usepackage{pgfplots}

\usepackage{soul}
\usepackage{colortbl} %
\usepackage{pifont} %

\pgfplotsset{compat=1.17}

\providecommand{\BibTeX}{{\rm B\kern-.05em{\sc i\kern-.025em b}\kern-.08em\TeX}}

\newcommand{\cmark}{\textcolor{barfix}{\ding{51}}}
\newcommand{\xmark}{\textcolor{barbare}{\ding{55}}}

\newcommand{\sys}{\textsc{Beaver}}

\newcommand{\uplink}{uplink}

\definecolor{barsolo}{HTML}{4E7FA3}
\definecolor{barbare}{HTML}{C0504D}
\definecolor{barfix}{HTML}{3A7A4F}
\definecolor{barpoll}{HTML}{B97A2A}
\definecolor{barqos}{HTML}{3A7A4F}
\definecolor{rescompute}{HTML}{8E6FA8}
\definecolor{resbw}{HTML}{5590BE}
\definecolor{grouprow}{HTML}{F1F3F6}
\definecolor{oursrow}{HTML}{EDF4EF} %

\newif\ifshowcomments
\showcommentstrue

\input{macro}

\begin{document}

\setlength{\abovedisplayskip}{3pt}
\setlength{\belowdisplayskip}{3pt}
\setlength{\abovedisplayshortskip}{1pt}
\setlength{\belowdisplayshortskip}{1pt}

\settopmatter{printacmref=false}

\title{{\sys}: Elastic GPU Sharing between ML and Latency-Critical vRAN Workloads}

\author{Yuncheng Yao$^{1}$, Zhenzhou Qi$^{2}$, Junyao Zheng$^{2}$, Chung-Hsuan Tung$^{2}$,\\Danyang Zhuo$^{1}$, Tingjun Chen$^{2}$}
\affiliation{%
  \institution{Departments of Computer Science$^{1}$ and Electrical \& Computer Engineering$^{2}$, Duke University}
  \city{}
  \state{}
  \country{}
}
\renewcommand{\shortauthors}{}

\input{sections/00_abstract}

\maketitle
\pagestyle{plain}

\input{sections/01_introduction}

\input{sections/02_background}

\input{sections/03_design}

\input{sections/05_implementation}

\input{sections/06_evaluation}
\input{sections/06b_dynamic_partition_eval}

\input{sections/06c_abalation}

\input{sections/07_realted_work}

\input{sections/08_conclusion}

\begin{acks}
This work was supported in part by NSF under grants CNS-2112562, CNS-2211944, ECCS-2434131, CNS-2443137, CNS-2450567, and OAC-2503010, NVIDIA Academic Grants, and the NVIDIA/Supermicro Grace Enablement Seeding Program.
This work was also supported in part by the Center for Ubiquitous Connectivity (CUbiC), sponsored by Semiconductor Research Corporation (SRC) and Defense Advanced Research Projects Agency (DARPA) under the JUMP 2.0 program.
\end{acks}

\bibliographystyle{ACM-Reference-Format}
\bibliography{refs}

\appendix
\input{sections/09_appendix}

\end{document}

%% file: macro.tex
\definecolor{lightgray}{gray}{0.9}
\definecolor{lightblue}{rgb}{0.9,0.9,1}
\definecolor{LightMagenta}{rgb}{1,0.5,1}
\definecolor{red}{rgb}{1,0,0}
\definecolor{brightgreen}{rgb}{0.4, 1.0, 0.0}

\newcommand{\remove}[1]{}

\newcommand{\hlc}[2][yellow]{ {\sethlcolor{#1} \hl{#2}} }

\ifshowcomments
  \newcommand\note[1]{\hlc[SkyBlue]{#1}}
  \newcommand\mynote[1]{\hlc[yellow]{#1}}
  \newcommand\tingjun[1]{\hlc[yellow]{TC: #1}}
  
  \newcommand\peter[1]{\hlc[LightMagenta]{Peter: #1}}
  \newcommand\tom[1]{\hlc[brightgreen]{TOM: #1}}
  \newcommand\andy[1]{\hlc[pink]{Andy: #1}}
  \newcommand\danyang[1]{\hlc[blue]{Danyang: #1}}
\else
  \newcommand\note[1]{}
  \newcommand\mynote[1]{}
  \newcommand\tingjun[1]{}
  \newcommand\peter[1]{}
  \newcommand\tom[1]{}
  \newcommand\andy[1]{}
  \newcommand\danyang[1]{}
\fi

\newcommand{\myparatight}[1]{\vspace{0.5ex}\noindent\textbf{#1~~}}

\newcommand*\circled[1]{\tikz[baseline=(char.base)]{
            \node[shape=circle,draw,fill=black,text=white,inner sep=0.6pt] (char) {\scriptsize{#1}};}}

\newcommand{\specialcell}[2][c]{%
  \begin{tabular}[#1]{@{}c@{}}#2\end{tabular}}

\newcommand{\littlesum}{\mathop{\textstyle\sum}}

\newcommand{\myCodeShort}[1]{\texttt{\small{#1}}}

\newcommand{\us}{$\upmu$s} %

\newcolumntype{+}{>{\global\let\currentrowstyle\relax}}
\newcolumntype{^}{>{\currentrowstyle}}

%% file: sections/00_abstract.tex
\begin{abstract}
Within the shared industry vision of AI-RAN, AI-and-RAN seeks to co-locate virtualized radio access network (vRAN) workloads and AI services on shared GPUs.
This sharing is inherently asymmetric: vRAN workload is latency-critical, whereas the machine learning (ML) workload is a throughput-oriented, best-effort co-tenant.
We present {\sys}, a GPU sharing system that jointly manages compute and memory resources while protecting the vRAN's strict processing deadline.
{\sys} sizes the vRAN's streaming multiprocessor (SM) allocation from each slot's scheduled workload, repartitions SM allocations at slot granularity, and rewrites compiled ML kernels to yield HBM bandwidth during the vRAN's memory-critical phases.
We implement {\sys} and evaluate it using NVIDIA Aerial with heterogeneous multi-cell workloads, real-world cellular traces, production inference kernels, and full-stack LLM serving.
On an H200 GPU, {\sys} keeps the vRAN's p99.9 latency within its 1.5\,{ms} uplink deadline while retaining 74\% of Llama-3.3-70B serving throughput.
It also incurs no observed deadline misses under replayed cellular traces, protects a 375\,{\us} downlink deadline, and generalizes to other GPUs including A100, GB10 and GH200.

\end{abstract}

%% file: sections/01_introduction.tex
\section{Introduction}
Co-locating virtualized radio access network (vRAN) workloads and AI services on shared GPU infrastructure is becoming an industry-wide objective with the AI-RAN vision~\cite{airan_alliance_vision, nvidia_ai_aerial}.
NVIDIA Aerial makes GPU-based 5G layer-1 (L1) processing practical~\cite{nvidia_aerial}, motivating recent systems such as YinYangRAN~\cite{yinyangran} and RAN-LLM~\cite{llm_ran_cloud} to explore co-locating radio workloads (e.g., vRAN baseband processing) with ML workloads (e.g., LLM inference serving) on the same GPU.
This sharing is inherently \emph{asymmetric}: the latency-critical vRAN workload is the higher-priority tenant, whereas the AI service is a best-effort co-tenant that may use only the GPU resources that can be safely spared without violating the strict vRAN processing deadline.
The central challenge is therefore to guarantee vRAN's latency while preserving as much co-tenant throughput as possible. 

\begin{figure}[!t]
\centering
\includegraphics[width=0.97\columnwidth]{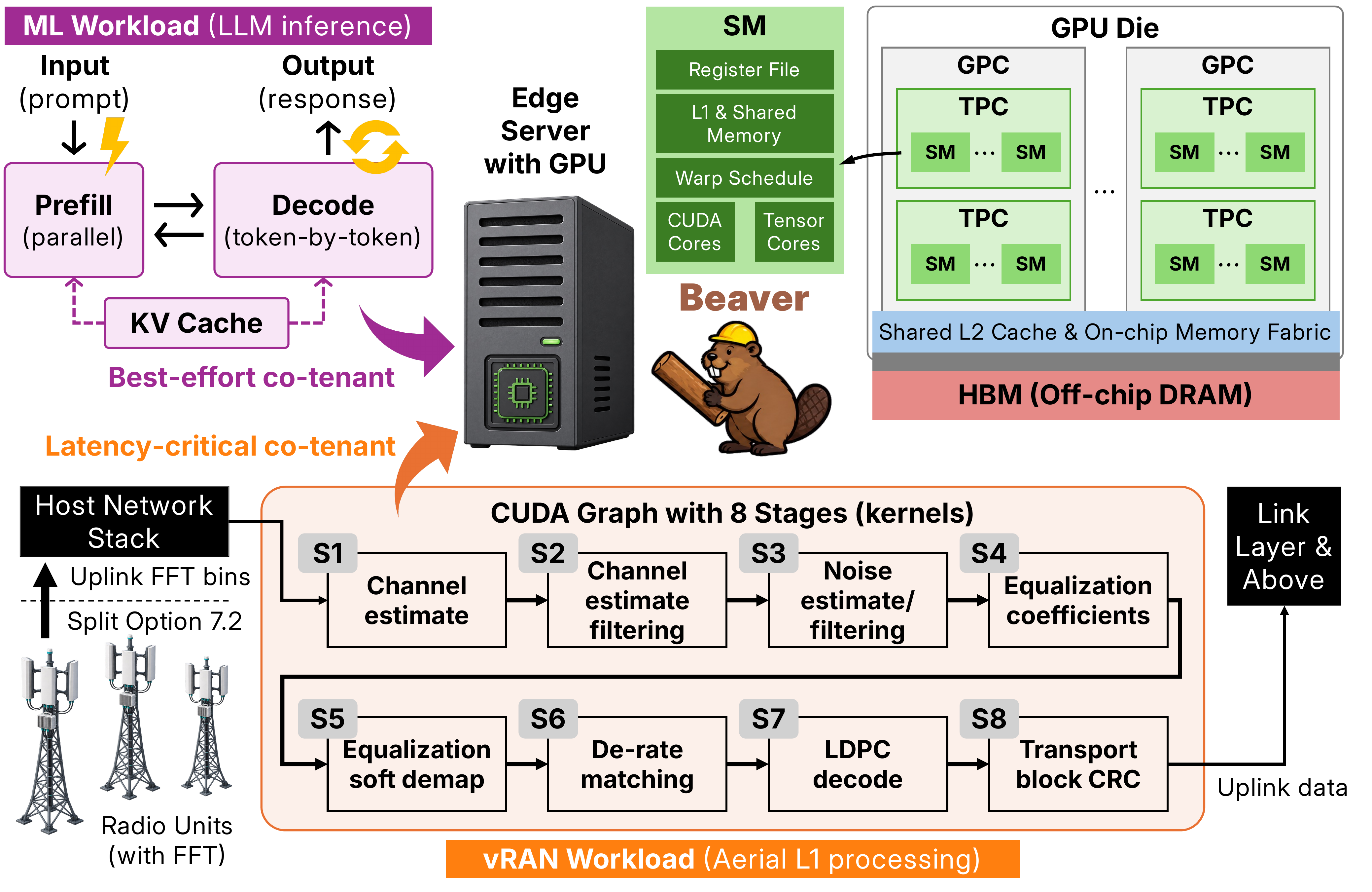}
\vspace{-3.5mm}
\caption{AI-and-RAN co-tenancy setting.
Radio units deliver uplink samples to a latency-critical Aerial L1 pipeline, while a best-effort LLM service shares the same GPU, whose SMs can be partitioned, but the L2 cache and HBM fabric remain shared.
{\sys} coordinates both compute and memory resources to protect the vRAN's strict processing deadline.}
\label{fig:beaver-intro}
\vspace{-3.0mm}
\end{figure}

Fig.~\ref{fig:beaver-intro} illustrates our target setting.
5G radio units (RUs) stream uplink samples over the fronthaul into an eight-stage Aerial L1 pipeline.
Under 5G Frequency Range~1 (FR1) numerology~1, each slot spans 500\,{\us}, and its uplink processing must complete within three slots, or 1{,}500\,{\us}. 
Simultaneously, a vLLM service performs LLM inference---including prefill and decode---on the same GPU.
The sharing opportunity arises from the substantial GPU capacity left idle by the latency-oriented vRAN pipeline.
Profiling the production Aerial uplink pipeline on an NVIDIA H200 GPU~\cite{nvidia_h200}, we find that a 20-cell load uses about 60\% of peak \emph{streaming multiprocessor (SM)} compute throughput and less than 10\% of peak \emph{high bandwidth memory (HBM)} bandwidth.

\begin{figure}[!t]
\centering
\includegraphics[width=0.96\columnwidth]{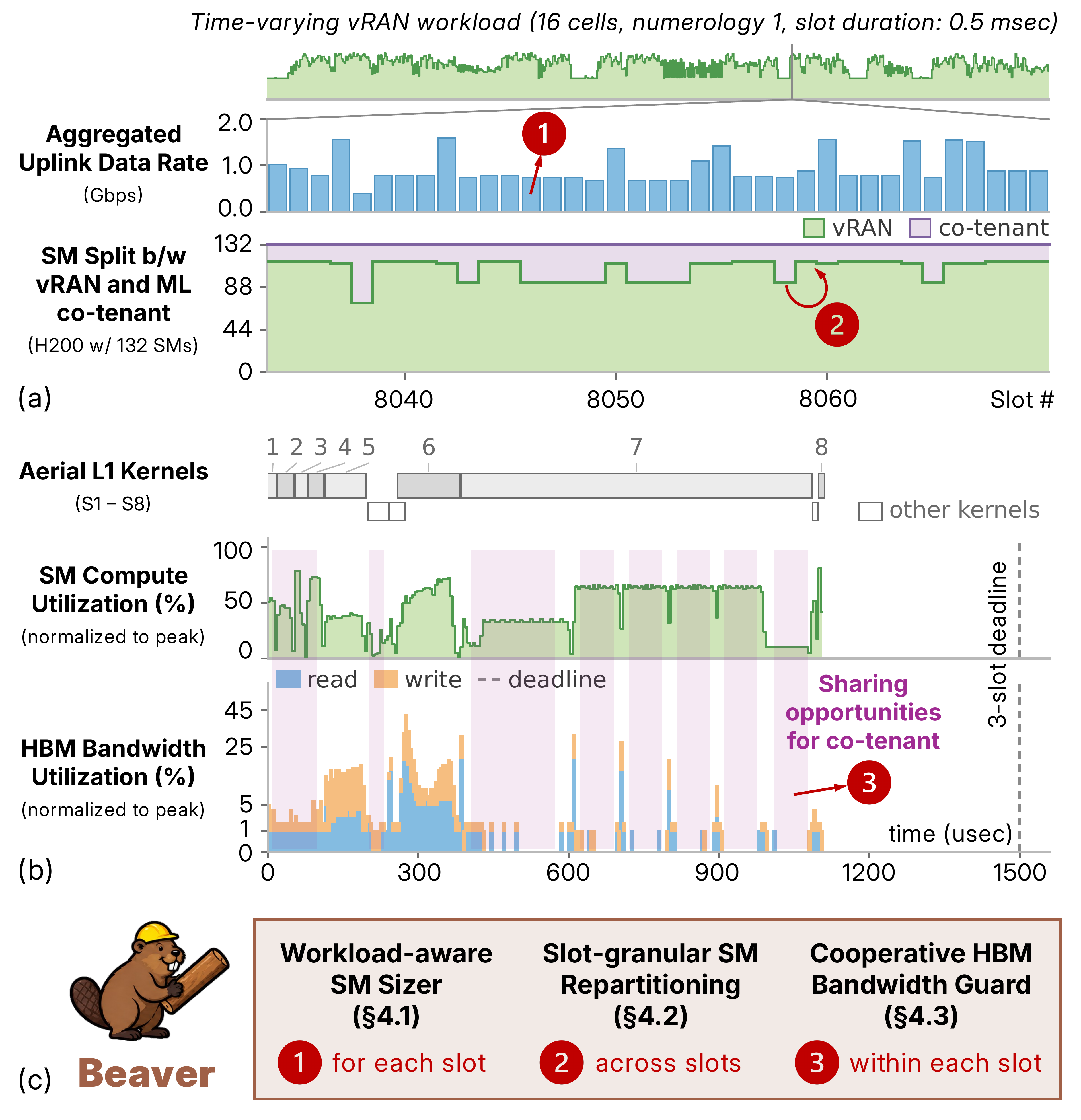}
\vspace{-3.5mm}
\caption{(a) Slot-varying vRAN uplink demand from a public cellular trace and the corresponding SM allocation selected by {\sys}.
(b) SM and HBM bandwidth utilization across the eight Aerial uplink stages (kernels) within one slot; shaded intervals indicate potential co-tenant sharing opportunities.
(c) {\sys}'s three modules and their operating timescale.}
\label{fig:beaver-overview}
\vspace{-3.0mm}
\end{figure}

This under-utilization is \emph{structural}.
To minimize latency, the vRAN pipeline executes a multi-stage CUDA graph comprising small, sequential kernels whose limited grid sizes and inter-stage dependencies cannot fully utilize the GPU; batching more work would improve utilization but increase latency.
Fig.~\ref{fig:beaver-overview} shows the resulting SM and HBM bandwidth headroom, both within individual slots and across consecutive slots under dynamic vRAN demand derived from a public cellular trace.
The shaded regions represent capacity that a co-tenant could potentially reclaim, provided that the vRAN's 99.9th-percentile (p99.9) uplink processing latency remains within its strict three-slot deadline.

Despite this opportunity, existing GPU sharing systems cannot simultaneously preserve the latency-critical vRAN deadline and reclaim spare capacity efficiently.
Recent GPU sharing designs for vRAN co-tenancy~\cite{yinyangran, llm_ran_cloud} rely primarily on spatial GPU partitions through NVIDIA multi-instance GPU (MIG)~\cite{nvidia_mig} or multi-process service (MPS)~\cite{nvidia_mps} with two key limitations.
First, SM partitioning isolates compute resources but not the shared L2 cache and HBM fabric (Fig.~\ref{fig:beaver-intro}). 
Consequently, although disjoint SMs eliminate compute contention, a bandwidth-intensive ML co-tenant (e.g., LLM decoding) can still contend with and delay vRAN's memory-bound kernels.
Second, existing SM partitions are mostly static and cannot be reconfigured at the radio slot timescale of 500\,{\us}.
The required SM budget varies from slot to slot (Fig.~\ref{fig:beaver-overview}(a))~\cite{perez2023madrid, groen2024tractor}, whereas reconfiguring MIG or MPS takes milliseconds to seconds (Table~\ref{tab:comparison-related-work}, \S\ref{ssec:eval-abalation}).

Achieving elastic, efficient, and safe GPU sharing between vRAN and ML workloads raises three challenges.
First, the system must determine, for dynamic and heterogeneous multi-cell workloads, the minimum vRAN SM budget required to meet the strict processing deadline.
Second, it must repartition the SMs between the vRAN and its co-tenant at sub-slot timescales.
Third, it must regulate the HBM bandwidth demand of opaque, compiled ML kernels without access to their source code or disrupting their synchronization.

We present {\sys}\footnote{Beavers build dams to regulate a shared water flow, mirroring how our system regulates GPU compute and memory resources to protect latency-critical vRAN while supporting ML co-tenancy.}, a GPU sharing system that addresses these challenges through three mechanisms operating at complementary timescales (Fig.~\ref{fig:beaver-overview}(c)).
\circled{1}~For each slot, a \emph{workload-aware SM sizer} predicts the smallest SM allocation that can complete the scheduled workload before its deadline  (\S\ref{ssec:design-latency-model}).
\circled{2}~Across slots, \emph{slot-granular SM repartitioning} applies the selected SM allocation by switching the vRAN to the desired CUDA green context~\cite{nvidia_greenctx} within only microseconds (\S\ref{sec:greenctx}).
\circled{3} Within each slot, a \emph{cooperative HBM bandwidth guard} rewrites the co-tenant's compiled kernels at the parallel thread execution (PTX) level, without requiring their source code, to poll a shared flag and back off whenever the vRAN workload signals a memory-critical phase.
HBM access is therefore coordinated between the workloads over time rather than divided into fixed bandwidth slices (\S\ref{sec:qos}).

We implement {\sys} as a GPU sharing layer comprising a runtime that interfaces with Aerial to size and switch SM partitions at slot granularity, and a compile-time rewriter that instruments both tenants' kernels at the PTX level (\S\ref{sec:impl}).
We evaluate {\sys} on an NVIDIA H200 using the Aerial uplink pipeline (\S\ref{sec:eval}).
Co-tenants include single-axis stressors; production inference kernels such as general matrix multiply (GEMM), FlashAttention-3, and fused mixture of experts (MoE); and a full-stack vLLM~\cite{kwon2023vllm} service running Llama-3.3-70B and Mistral-Small-24B.
With 13 memory-heavy vRAN cells and LLM serving co-tenants,
{\sys} keeps vRAN's p99.9 latency within the deadline while retaining 74\% of the Llama-3.3-70B serving throughput.
Under two real-world cellular traces~\cite{perez2023madrid, groen2024tractor} capturing changes in cell configuration and traffic intensity, {\sys} adapts the SM allocation every slot, incurs no deadline misses, and sustains LLM serving.
{\sys} also eliminates observed deadline misses with co-located production inference kernels, protects a 375\,{\us} downlink processing deadline, and generalizes to A100, GB10 and GH200 GPUs while retaining 78\%-85\% of vLLM throughput.

In summary, this paper makes the following contributions:
\begin{itemize}[leftmargin=*, topsep=2pt, itemsep=1pt]
  \item
  We \emph{identify and characterize} the GPU sharing opportunities created by the structurally under-utilized SM and HBM resources of the production NVIDIA Aerial vRAN pipeline;
  \item
  We \emph{design and implement} {\sys}, a GPU sharing system tailored to asymmetric co-tenancy between latency-critical vRAN and best-effort ML workloads, through a workload-aware SM sizer, slot-granular SM repartitioning, and a cooperative HBM bandwidth guard;
  \item
  We \emph{extensively evaluate} {\sys} using Aerial's uplink and downlink pipelines, heterogeneous multi-cell workloads, and real-world cellular traces, covering co-tenancy ML co-tenants ranging from production inference kernels to full-stack LLM serving, and across GPU architectures.
\end{itemize}
\emph{To the best of our knowledge, {\sys} is the first system that enables elastic GPU sharing between latency-critical vRAN and best-effort ML workloads, and demonstrates that a full LLM serving stack can share the GPU without violating the vRAN's strict processing deadline while preserving substantial co-tenant serving throughput.}
We will open-source {\sys}'s codebase and benchmarking suite.

\newcommand{\qual}[1]{{\scriptsize\textcolor{black!55}{#1}}}
\begin{table*}[t!]
\caption{Comparison of GPU sharing systems ({\cmark}: provided; {\xmark}: absent/none; SLO: service level objective).
\emph{Deadline met} names the timing guarantee each system makes and its scale; only {\sys} enforces a hard per-slot latency deadline.}
\label{tab:comparison-related-work}
\vspace{-3.0mm}
\centering\scriptsize
\setlength{\tabcolsep}{2.6pt}
\renewcommand{\arraystretch}{1.08}
\begin{tabular}{@{}>{\raggedright\arraybackslash}p{3.15cm} l l cc >{\raggedright\arraybackslash}p{2.0cm} cc l@{}}
\toprule
Approach & Protected tenant & Co-tenant evaluated & SM & HBM BW & Deadline met & Adaptive & Opaque Kernel & Reconfiguration \\
\midrule
\rowcolor{grouprow} \multicolumn{9}{@{}l}
{\textbf{Software GPU Sharing Systems}} \\
\textcolor{barsolo}{\textbf{Gandiva, AntMan, HiveD}} \qual{job scheduling}~\cite{xiao2018gandiva,antman2020,hived2020} & Guaranteed jobs & Opportunistic jobs & \xmark & \xmark & \xmark & \cmark & \cmark & seconds \qual{migrate} \\
\textcolor{barsolo}{\textbf{Salus, Orion, TGS}}~\cite{yu2020salus,strati2024orion,wu2023tgs} & High priority jobs & Best-effort jobs & \xmark & \xmark & Soft SLO \qual{10\,ms--1\,s} & \cmark & \cmark & ms \qual{per kernel/op} \\
\textcolor{barsolo}{\textbf{REEF}} \qual{kernel preemption}~\cite{han2022reef} & Real-time inference & Best-effort DNN & \xmark & \xmark & Preempt \qual{AMD~only} & \cmark & \xmark & $\mu$s \qual{preempt} \\
\rowcolor{grouprow} \multicolumn{9}{@{}l}{\textbf{GPU Sharing for vRANs}} \\
\textcolor{barsolo}{\textbf{CAORA}}~\cite{caora} & 3 BS traces & None & \cmark & Fixed MIG slice & \xmark & \xmark & \cmark & seconds \qual{MIG rebuild} \\
\textcolor{barsolo}{\textbf{RAN-LLM}}~\cite{llm_ran_cloud} & 1 cell & LLMs \qual{up to 7B} & \cmark & Fixed MIG slice & \xmark & \xmark & \cmark & seconds \qual{MIG rebuild} \\
\textcolor{barsolo}{\textbf{YinYangRAN}}~\cite{yinyangran} & 4 cells & RNN \qual{up to 1M} & \cmark & \xmark & Soft SLO \qual{$\sim$1\,ms} & \xmark & \cmark & $\sim$564\,ms \qual{MPS restart} \\
\rowcolor{oursrow} \textcolor{barfix}{\textbf{{\sys} (This Work)}} & \textbf{Hetero multi-cell} & \textbf{LLM} \qual{up to 70B} & \cmark & \textbf{Phase-aware control} & \textbf{Hard, 1500\us} & \cmark & \cmark & \textbf{$\sim$1.8\us} \qual{pointer swap} \\
\bottomrule
\end{tabular}
\end{table*}

%% file: sections/02_background.tex
\section{Background}
\label{sec:background}

\myparatight{5G vRAN baseband and GPU-accelerated L1.}
The physical layer (L1) converts coded transport blocks into downlink radio symbols and recovers them from uplink symbols.
These operations traditionally run on dedicated FPGAs or ASICs; software-defined basebands use commodity CPUs and GPUs, enabling flexible updates and compute allocation as configurations and load change.
NVIDIA Aerial implements 5G L1 using GPU libraries and CUDA kernels~\cite{nvidia_aerial}.
A kernel comprises thread blocks assigned to SMs~\cite{cuda_programming_guide}.
Multiple blocks can share an SM when resources permit; L1 latency therefore depends on SM occupancy, memory traffic, and specialized execution units.
Concurrent ML kernels can contend for these resources even across different CUDA streams.

\myparatight{Uplink/downlink processing and hard deadline.}
In 5G, the physical downlink shared channel (PDSCH) carries user data from the base station to a user equipment (UE).
Its transmit pipeline performs cyclic redundancy check (CRC) attachment, low-density parity-check (LDPC) encoding, rate matching, modulation, MIMO precoding, resource mapping, and orthogonal frequency-division multiplexing (OFDM) waveform generation.
Conversely, the physical uplink shared channel (PUSCH) receive pipeline transforms incoming I/Q samples, estimates channel and noise, equalizes MIMO layers, produces log-likelihood ratios (LLRs), and performs de-rate matching, LDPC decoding, and CRC checking~\cite{tgpp_38211}.
PUSCH is generally more \emph{compute-intensive}: reception compensates for a time-varying channel, and LDPC decoding may require multiple iterations.
We therefore use PUSCH as the representative protected vRAN workload.
Fig.~\ref{fig:beaver-intro} shows the eight stages of NVIDIA's Aerial L1 uplink processing, where missing a slot's deadline can trigger retransmission and consume additional air-interface capacity; safe GPU sharing must therefore protect tail latency.
Following 3GPP timing constraints~\cite{tgpp_38214}, we target a 1.5\,ms uplink deadline.

\myparatight{GPU resources: SM and HBM.}
A GPU groups execution into \emph{SMs}~\cite{nvidia_h200} (see Fig.~\ref{fig:beaver-intro}), and places kernel blocks on available SMs in successive \emph{waves} when they cannot all reside concurrently; a running block remains on its SM until completion~\cite{cuda_programming_guide,gilman2021tbs}, and a kernel waits when all SMs are full.
Disjoint SM partitions prevent both HOL blocking and compute contention but still share the L2 cache and HBM system.
Because HBM bandwidth is shared, LLM decoding that streams weights from HBM can delay the radio through memory bandwidth contention.
NVIDIA exposes three relevant partitioning mechanisms. 
\textbf{MIG}~\cite{nvidia_mig} isolates SMs and proportional memory slices, but it takes \emph{seconds} to rebuild.
\textbf{MPS}~\cite{nvidia_mps} caps active thread capacity, while \textbf{CUDA GC}~(green context) assigns disjoint SMs within one process~\cite{nvidia_greenctx}; both leave HBM shared and cannot resize on the slot clock.
Compute-only sharing of this kind is common practice in software GPU sharing systems built for ML-only workloads (Table~\ref{tab:comparison-related-work}).
However, none combines slot-granular adaptation with HBM protection.

\myparatight{GPU compilation pipeline.}
During compilation, CUDA code is first lowered to PTX, NVIDIA's virtual instruction set, then assembled into the GPU's architecture-specific native machine code in streaming assembly (SASS) as a \emph{cubin}~\cite{cuda_programming_guide}.
Libraries may bundle PTX with cubins, while runtime generators expose an interceptable just-in-time (JIT) path; PTX can thus come from a fat binary or JIT.

%% file: sections/03_design.tex
\section{Sharing Opportunities and Challenges}
\label{sec:motivation}

\begin{figure}[!t]
\centering
\includegraphics[width=0.98\columnwidth]{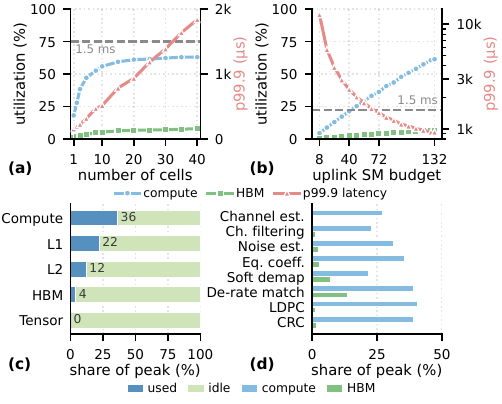}
\vspace{-3.0mm}
\caption{GPU headroom for memory-intensive uplink processing without a co-tenant; percentages are normalized to the full H200 GPU with 132 SMs. 
Utilization and p99.9 latency versus (a) cell count and (b) SM budget for 20 homogeneous cells; 72 SMs is the smallest deadline-safe allocation.
(c) Resource utilization and (d) per-kernel compute and HBM utilization of the 20-cell workload with 72 SMs.}
\label{fig:motivation-gpu-headroom}
\vspace{-3.0mm}
\end{figure}

\subsection{GPU Utilization of vRAN Workloads}

We consider a set of homogeneous \emph{memory-heavy} cells, each with 273 PRBs, four MIMO layers, and MCS-27 (test vector \myCodeShort{tv-28}), running the NVIDIA Aerial~\cite{nvidia_aerial} L1 processing on an H200 GPU featuring 132 SMs and 141\,GB of HBM3e memory with 4.8\,TB/s memory bandwidth.
For detailed experimental setup and test vectors, see \S\ref{sec:impl} and Appendix~\ref{sec:appendix-tv}.

Fig.~\ref{fig:motivation-gpu-headroom}(a) reports the SM compute utilization, HBM bandwidth utilization, and p99.9 uplink latency on an unpartitioned H200 (i.e., using all 132 SMs) as the number of concurrently served cells increases.
It can be seen that the p99.9 uplink latency reaches the 1.5\,ms deadline at 32 cells even though neither resource is saturated: compute utilization rises quickly but plateaus near 60\% beyond 15 cells, while HBM bandwidth utilization remains below 10\% even at 40 cells.
Fig.~\ref{fig:motivation-gpu-headroom}(b) fixes the 20-cell vRAN workload and varies the number of SMs allocated to it.
Reducing the allocation increases tail latency, and 72 SMs is the smallest allocation that satisfies the p99.9 uplink deadline for 20 cells, leaving 60 SMs for a co-tenant. 
The detailed resource breakdown for this 20-cell workload with 72 SMs is shown in Fig.~\ref{fig:motivation-gpu-headroom}(c), normalized to the capacity of the entire GPU, where compute, L1/L2 caches, and HBM bandwidth all remain well below their respective limits, with no tensor core usage.
Fig.~\ref{fig:motivation-gpu-headroom}(d) further shows the compute and HBM utilization of individual vRAN kernels: no kernel approaches peak compute throughput, and none draws more than a quarter of peak HBM bandwidth.
The sequential dependencies within each cell’s processing pipeline therefore leave a substantial fraction of the GPU idle, creating headroom for a co-tenant while guaranteeing that the vRAN workload can meet its tight latency deadline.

\subsection{Resource Contention in GPU Sharing}

\begin{figure}[!t]
\centering
\includegraphics[width=0.90\columnwidth]{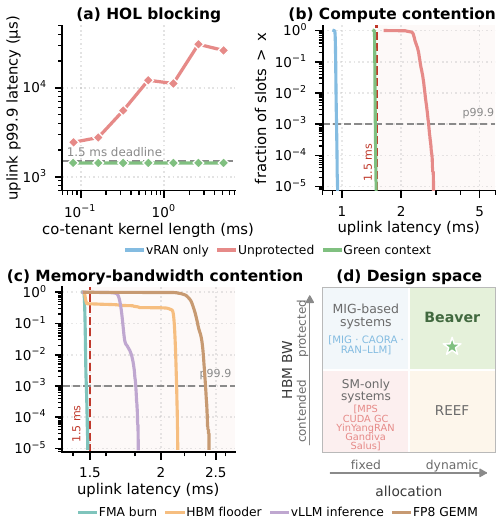}
\vspace{-3.0mm}
\caption{GPU sharing failure modes and design space.
(a) HOL blocking couples tail latency to co-tenant kernel length.
(b) Shared SMs cause compute contention.
(c) Disjoint SMs eliminate compute contention but not HBM contention.
(d) {\sys} explores dynamic SM repartitioning while protecting the HBM bandwidth between co-tenants.}
\label{fig:motivation-failures}
\vspace{-3.0mm}
\end{figure}

After characterizing GPU utilization when the vRAN workload runs \emph{alone}, we examine what happens when it shares the GPU with another workload.
Because the vRAN must satisfy a per-slot latency deadline, we treat it as the first-class tenant and the co-located workload as a best-effort tenant.
Co-tenancy introduces three distinct effects:

\myparatight{HOL blocking and compute contention.}
A running CUDA block remains on its assigned SMs until it completes (see
\S\ref{sec:background}).
If a co-tenant occupies all SMs when a vRAN kernel is launched, the vRAN kernel must wait for some of those blocks to finish.
We refer to this delay as \emph{head-of-line (HOL) blocking}.
To isolate its effect, we use a compute stressor that repeatedly performs arithmetic.
Fig.~\ref{fig:motivation-failures}(a) plots the p99.9 uplink latency of the 20-cell workload as the compute stressor's kernel length increases, with and without SM partitioning.
Without SM partitioning (\emph{Unprotected}), the p99.9 latency increases with the stressor's kernel length since the vRAN workload queues behind its running blocks.
A 72:60 SM partitioning between the vRAN and stressor co-tenant keeps the uplink within its deadline by reserving dedicated SMs.
Even when the vRAN workload can begin immediately, blocks from the two tenants may execute concurrently on the same SMs and compete for warp slots, execution units, registers, and other resources.
We refer to this interference as \emph{compute contention}; it arises under unprotected execution (no SM partitioning) and is eliminated when the tenants receive disjoint SM partitions, e.g., via green contexts (GC)~\cite{nvidia_greenctx}.
Fig.~\ref{fig:motivation-failures}(b) plots the complementary CDF of uplink latency for the 20-cell workload co-running with the same compute stressor.
It compares unprotected execution, a 72:60 SM partition using GC, and the vRAN only baseline with no co-tenant.
HOL blocking mainly enlarges the tail by delaying when a vRAN kernel can start, whereas compute contention slows the kernel after it starts and consequently affects most vRAN slots.

\myparatight{HBM bandwidth contention.} 
In addition to compute interference, contention in the shared memory hierarchy presents a separate limitation and requires a different mechanism, as assigning the co-tenants disjoint SMs (e.g., via CUDA GC) does not partition the GPU's caches or HBM fabric.
Therefore, a memory-intensive co-tenant can still delay the vRAN workload even when both workloads never execute on the same SM.
Fig.~\ref{fig:motivation-failures}(c) plots the same complementary CDF for the 20-cell workload under a 72:60 SM split across four distinct co-tenant workloads (see \S\ref{sec:impl}); a curve that crosses the horizontal p99.9 line misses the vRAN deadline.
The compute stressor causes no deadline misses, confirming that the disjoint SM allocation removes compute contention.
In contrast, HBM flooder, FP8 General Matrix Multiplications (GEMM), and vLLM inference co-tenants continue to contend with vRAN for the shared HBM fabric, producing significant deadline miss rates of up to 99.8\%.

\subsection{Design Challenges}
\label{ssec:motivation-challenges}
 
Safe and efficient vRAN co-tenancy requires both dynamic SM allocation and protection from contended HBM bandwidth, posing the following design challenges (see Fig.~\ref{fig:motivation-failures}(d)).

\myparatight{\circled{C1} Partitioning SMs for heterogeneous vRAN workloads.}
Cell count alone cannot determine the vRAN's compute demand: heterogeneous cells require different SM budgets. 
Mapping each slot's workload to a deadline-safe SM partition is challenging because GPU performance does not scale linearly with the allocated SM count, and kernels differ in grid size, occupancy, synchronization, and memory behavior.
On Hopper GPUs, GC exposes partitions in increments of eight SMs, and {\sys} must map dynamic workloads onto these discrete choices and select the smallest SM partition that meets the vRAN deadline.
Allocating too few SMs causes deadline misses, while allocating too many unnecessarily reduces the resources available to the co-tenant.

\myparatight{\circled{C2} Adaptive SM partition at slot granularity.}
vRAN workload can change substantially from one slot to the next at the sub-millisecond level, making a static partition either unsafe under peak load or wasteful under light load.
{\sys} must therefore resize the vRAN partition at per-slot granularity while ensuring that every slot meets its hard deadline.
However, existing GPU partitioning mechanisms such as MIG and MPS are not designed for such rapid adaptation (see \S\ref{sec:background}).
GC provides SM isolation, but the SM assignments are fixed when the context is created and cannot be resized in place. 
Changing a partition therefore requires creating or switching contexts and waiting for work in the old partition to drain before its SMs can be reclaimed, making direct reconfiguration too expensive for sub-millisecond control.

\myparatight{\circled{C3} Controlling opaque ML co-tenants.}
On the other hand, ML workloads typically invoke kernels from vendor libraries, packaged frameworks, and runtime generators (e.g., Triton).
These kernels are \emph{opaque} to the application and often employ asynchronous copies and warp-specialized synchronization, making source modification impractical and potentially unsafe. 
Since SM partitioning alone---even through the use of GC---does not isolate the shared HBM subsystem, the vRAN workload's memory demand can vary sharply across slots, while concurrent ML kernels can saturate HBM bandwidth and queues during these critical phases.
Once issued, these memory requests cannot be recalled, so coarse-grained or reactive throttling cannot protect vRAN's tail latency. 
{\sys} must therefore regulate ML co-tenant's execution transparently at sub-slot granularity, preserve kernel semantics, and protect the vRAN deadline without unnecessarily reducing ML co-tenant's throughput.

\section{{\sys}'s Design}
\label{sec:design}

{\sys} addresses these challenges by combining deadline-driven vRAN resource provisioning with low-overhead SM repartitioning and cooperative control of co-tenant memory traffic.
We focus on uplink processing because its iterative LDPC decoding makes it substantially more intensive than downlink processing (see \S\ref{sec:background}).
{\sys} comprises three functional modules, illustrated in Fig.~\ref{fig:beaver-overview} and described below.

\subsection{Workload-Aware SM Sizer}
\label{ssec:design-latency-model}

\input{sections/04b_partition_controller}

\subsection{Slot-Granular SM Repartitioning}
\label{sec:greenctx}

\begin{figure}[!t]
\centering
\includegraphics[width=0.95\columnwidth]{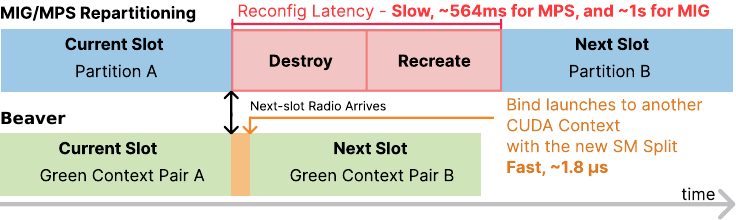}
\vspace{-3.0mm}
\caption{{\sys} enables slot-granular SM repartitioning (in only 1.8\,{\us}) by rebinding launches to a pre-created GC pair.}
\label{fig:gc-reconfiguration}
\vspace{-3.0mm}
\end{figure}

{\sys} must adapt the SM allocation at slot granularity (0.5\,ms in numerology~1) without violating the processing deadline, which cannot be achieved by existing SM partition mechanisms such as MIG and MPS (see \S\ref{ssec:motivation-challenges} and Fig.~\ref{fig:gc-reconfiguration}).
CUDA GC provides finer-grained SM isolation, but its resources are immutable after creation~\cite{nvidia_greenctx}; creating one on H200 takes $\approx$1.6\,ms, already exceeding the slot duration.

\begin{figure*}[!t]
\centering
\includegraphics[width=0.95\textwidth]{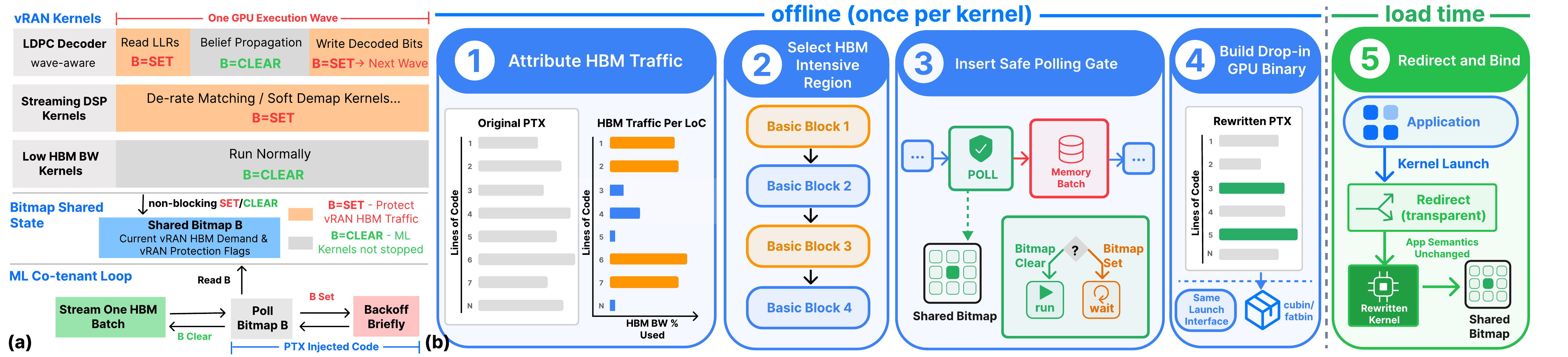}
\vspace{-3.0mm}
\caption{{\sys}'s cooperative HBM bandwidth guard:
(a) How each kernel's
bitmap signaling schedule follows its HBM-bandwidth pattern; the ML co-tenant
polls the shared bitmap and backs off while any flag is set.
(b) Profile-guided PTX injection: the offline pass that finds the
HBM-heavy regions and rewrites them, and the load-time redirect that binds the
shared bitmap.}
\label{fig:qos-pipeline}
\vspace{-2.0mm}
\end{figure*}

{\sys} addresses this mismatch with a \emph{slot-granular SM repartitioning} module that separates slow SM partition construction from fast runtime selection, as shown in Fig.~\ref{fig:gc-reconfiguration}(\emph{bottom}).
At startup, it pre-materializes a pool of complementary GC pairs, one context for the vRAN and one for the ML co-tenant.
Each GC pair represents a supported split $s$:$(M-s)$, where $M$ is the GPU's total SM count and $s$ is the vRAN allocation.
For example, an H200 contains 132 SMs and GC interface changes the vRAN allocation in increments of eight SMs.
{\sys} therefore constructs the 17 candidate partitions:
\begin{align}
    \mathcal{P}_{\text{H200}} = \left\{ (8j :132-8j); \forall j=0,\dots,16 \right\},
    \label{eq:h200-sm-partition}
\end{align}
which correspond to \{ 0:132, 8:124, $\dots$, 128:4 \} SM partitions across the vRAN and co-tenant workloads.
Pre-materializing the complete hardware-supported action space decouples the runtime controller from context-creation latency.

Before each slot, {\sys}'s SM sizer computes the minimum vRAN SM count required to meet the deadline. 
Since the requested count may not coincide with a supported split, it will be rounded up to the smallest $s$ represented in $\mathcal{P}_{\text{H200}}$. 
It then launches the next slot's vRAN and co-tenant workloads in their corresponding GC pair.
Specifically, a switch changes the contexts used by subsequent kernel launches; it does not migrate blocks that are already executing.
{\sys} therefore performs the selection at the slot boundary and directs new work to the selected pair, while previously launched work completes in its original context.
Because both contexts already exist, the critical-path operation is a launch-path rebind rather than context construction.
This rebind takes only $\approx$1.8\,{\us}, less than {0.5\%} of the 0.5\,ms slot duration, and nearly 900$\times$ less than the measured GC creation time.

\subsection{Cooperative HBM Bandwidth Guard}
\label{sec:qos}

SM partitioning does not partition the GPU's HBM subsystem or bandwidth.
A memory-intensive ML co-tenant can therefore delay the vRAN workload even when the two workloads execute on disjoint SMs.
{\sys} addresses this interference in the HBM domain with an \emph{asymmetric, cooperative HBM bandwidth guard} module: the \emph{latency-critical} vRAN workload announces short intervals in which its latency depends critically on HBM bandwidth, during which the \emph{best-effort} co-tenant defers new memory traffic.

\myparatight{Cooperative bitmap protocol.}
Fig.~\ref{fig:qos-pipeline}(a) shows the bitmap shared by the two tenants, $\mathbf{b} = [b_1, \dots, b_{N_b}]$.
Each bit represents one vRAN thread block, and $N_b$ is sized for the maximum number of thread blocks that can overlap in the vRAN pipeline.
The profile described below determines which parts of a kernel require protection, but each executing thread block signals independently.
Before entering an HBM-critical region, a thread block sets its bit; after leaving the region, it clears the same bit.
Protection remains active while any bit is set, i.e., $\bigvee_{j=1}^{N_b} b_j$.
Per-block bits are necessary because thread blocks progress independently: one block may leave its HBM-critical region while another is still issuing HBM requests.
A single kernel-wide bit could therefore be cleared too early, whereas the bitmap remains set until the last protected block clears its bit.
The bitmap occupies only a few machine words, and repeated ML polls normally hit in cache.
Before an instrumented co-tenant kernel begins another HBM-intensive region, it checks the bitmap:
If any bit is set, the participating threads briefly back off and poll again; otherwise, they proceed immediately.
This guard stops new co-tenant memory traffic but cannot preempt requests already in flight.
We therefore place polling gates frequently enough to bound the traffic issued after a vRAN thread block requests protection.

\begin{figure}[!t]
\centering
\includegraphics[width=0.98\linewidth]{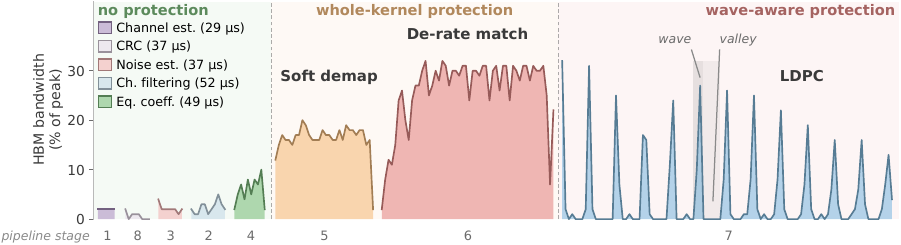}
\vspace{-3.0mm}
\caption{Measured per-kernel HBM demand across one uplink slot, normalized to the GPU's peak HBM bandwidth; stage numbers indicate kernel order (see also Fig.~\ref{fig:beaver-intro}).}
\label{fig:vran-hbm-profile}
\vspace{-3.0mm}
\end{figure}

\myparatight{Profile-guided bitmap signaling.}
Protecting every vRAN kernel for its entire execution would preserve the radio deadline but unnecessarily suppress ML progress.
Because the vRAN workload follows a fixed, slot-periodic pipeline, {\sys} profiles each kernel's time-resolved HBM demand offline and assigns it one of the three signaling schedules, as illustrated in Fig.~\ref{fig:vran-hbm-profile}.
First, kernels such as channel estimation, CRC, noise estimation, channel filtering, and equalization generate relatively little HBM traffic, therefore requiring \emph{no protection} and the ML co-tenant runs without interruption.
Second, streaming kernels such as soft demapping and rate de-matching sustain high HBM demand throughout execution.
They set their bit at kernel entry and clear it at kernel exit, thereby receiving \emph{whole-kernel protection}.
While the bit is set, the co-tenant defers new HBM-intensive regions; HBM-light computation may continue. 
Third, LDPC decoding exhibits a bursty, wave-structured traffic pattern.
As Fig.~\ref{fig:vran-hbm-profile} shows, each execution wave contains an HBM-intensive peak followed by an HBM-light valley.
A wave first reads LLRs from HBM, performs belief propagation with little HBM traffic, and then writes its decoded bits.
For such a kernel, whole-kernel protection would also block the ML tenant during the long belief propagation valleys. 
{\sys} instead applies \emph{wave-aware protection}: it sets the bitmap during each wave's reads and writes but clears it during belief propagation so that the co-tenant can make progress.

\myparatight{Injecting the guard without source.}
ML co-tenants commonly invoke vendor GEMM, attention, and projection kernels whose CUDA source is unavailable.
{\sys} therefore rewrites kernels with available PTX (see \S\ref{sec:background}) and substitutes the rewritten kernels at load time.
Fig.~\ref{fig:qos-pipeline} shows this five-stage pipeline: Stages~1--4 run once offline for each targeted kernel, while Stage~5 activates the rewritten kernel at load time.
The vRAN workload profiling determines \emph{when} protection is needed; this pipeline determines \emph{where} the ML co-tenant should pause before issuing more HBM traffic.
Specifically:
\begin{enumerate}[leftmargin=*, topsep=2pt, itemsep=1pt]
\item 
\underline{\emph{Attribute HBM traffic}}:
PTX identifies instructions that access memory, but does not show how often they execute or how much HBM traffic they generate.
{\sys} profiles the unmodified kernel and attributes its measured HBM traffic to individual PTX lines, and the results identify the code that contributes most to the kernel's HBM demand.
\item
\underline{\emph{Select HBM-intensive regions}}:
Polling before every memory instruction would add excessive overhead.
{\sys} instead groups PTX lines into \emph{basic blocks}---straight-line code regions with one entry and no internal branches~\cite{aho2006compilers}.
It sums the traffic profiled from Stage~1 within each block and selects the smallest set of blocks that covers a configured fraction of the kernel's HBM traffic.
As a result, a small set of useful candidate locations is identified, at each of which Stage~3 inserts a safe gate.
\item
\underline{\emph{Insert a safe polling gate}}:
{\sys} cannot use the same gate placement everywhere since GPU kernels organize their memory-producing threads differently.
We call a gate \emph{safe} when every thread that can issue the selected memory batch encounters it, without requiring unrelated threads to participate.
{\sys} classifies each selected region by its execution structure and chooses the corresponding gate.
If the shared bitmap is clear, the gated threads issue the next memory batch; otherwise, they briefly back off and check again.
\item
\underline{\emph{Build a drop-in GPU binary}}:
{\sys} reassembles the rewritten PTX into a GPU binary using the original target architecture and compiler settings.
The kernel retains the same arguments and launch interface.
\item
\underline{\emph{Redirect and bind}}:
At load time, a runtime shim redirects each targeted kernel to its rewritten binary and maps the shared bitmap into both tenants' address spaces.
The injected poll is thereby bound to the vRAN signal without changing the ML co-tenant's framework or application.
\end{enumerate}

%% file: sections/04b_partition_controller.tex
The first module models GPU-based multi-cell vRAN workload processing latency and determines the smallest vRAN SM allocation that can complete the vRAN workload before its slot deadline.
This is not a trivial task, as profiling every possible vRAN workload with multiple heterogeneous cells is prohibitively expensive:
Suppose the system supports $V$ cell configurations, or \emph{test vectors}, and may concurrently serve $C$ cells, there are $V^C$ possible workload profiles for every slot.
Our key idea is to profile only workloads with \emph{homogeneous} cells, in which all cells use the same configuration in a slot, and then compose fitted parameters to predict previously unseen workloads with \emph{heterogeneous} cells.

\myparatight{Processing latency modeling for homogeneous cells.}
Let $C$ denote the cell count and $M$ denote the total number of SMs on the GPU.
For a slot with homogeneous cells, we model the \emph{median} uplink processing latency as a function of $C$, the cell configuration $\bm{\uptheta}$ (test vector), and the number of SMs allocated to the vRAN workload $s$, given by
\begin{align}
& L_{\text{p50}} \left( C, \bm{\uptheta}, s \right)
= \tau_{0}(\bm{\uptheta}) 
+ C \cdot \tau_{1}(\bm{\uptheta})
+ \frac{M}{s} \cdot C \cdot \tau_{2}(\bm{\uptheta}).
\end{align}
This model consists of three interpretable coefficients:
\begin{itemize}[leftmargin=*, topsep=2pt, itemsep=1pt]
    \item 
    \emph{Setup time}, $\tau_{0}(\bm{\uptheta})$, represents the per-slot overhead (e.g., initiating the slot’s GPU kernel launches);
    \item
    \emph{SM-insensitive time}, $\tau_{1}(\bm{\uptheta})$, represents per-cell work whose latency changes little with the SM allocation because it exposes limited thread-block parallelism (e.g., small-grid kernels such as RSSI/RSRP measurement); 
    \item
    \emph{SM-parallel time}, $\tau_{2}(\bm{\uptheta})$, represents per-cell work whose latency decreases as more SMs are allocated (e.g., LDPC decoding and rate de-matching).
\end{itemize}
We fit the model offline using end-to-end measurements for every supported cell configuration $\bm{\uptheta} \in \bm{\upTheta}$, as specified by its MCS, payload size, and code block count~\cite{tgpp_38214} (Table~\ref{tab:tvintensity} in Appendix~\ref{sec:appendix-tv}).
For each supported $\bm{\uptheta}$, we vary the $(C, s) \in \mathcal{C} \times \mathcal{S}$ and record the median processing latency, $\widetilde{L}_{\text{p50}} \left( C, \bm{\uptheta}, s \right)$.
Here, $\mathcal{C}$ and $\mathcal{S}$ denote the set of possible cell counts and SM allocations.
A least-squares regression then jointly estimates
\begin{align}
& \big\{ \widehat{\tau}_0(\bm{\uptheta}), \widehat{\tau}_1(\bm{\uptheta}), \widehat{\tau}_2(\bm{\uptheta}) \big\} \nonumber \\
& = \arg\min\ \littlesum_{C \in \mathcal{C}} \littlesum_{s \in \mathcal{S}} \left[ \widetilde{L}_{\text{p50}}(C, \bm{\uptheta},s) - \tau_0 - C \cdot \tau_1 - \frac{M}{s} \cdot C \cdot \tau_2 \right]^2. \nonumber
\end{align}
Note that these coefficients are not obtained by separately measuring individual vRAN kernels or DSP stages.
Instead, they are jointly fitted from end-to-end uplink latency measurements.
The terminology above provides an interpretation of the scaling behavior captured by each coefficient; a single kernel may affect more than one fitted coefficient.
Using this method, the profiling cost is linear in $V$ and $C$ rather than exponential.
Our implementation is calibrated on Aerial's 28 uplink test vectors, which span $V = |\bm{\upTheta}| = 28$ distinct radio configurations; Appendix~\ref{sec:appendix-sizing-terms} gives the profiling grid, the fitting procedure, and the fitted coefficients.

\myparatight{Processing latency prediction for heterogeneous cells.}
A slot may contain cells with different configurations at \emph{runtime}.
We compose the coefficients learned from homogeneous cell workloads and add a safety reserve to predict its p99.9 processing latency, given by
\begin{align}
& \widehat{L}_{\text{p99.9}} \left( \mathcal{C}(t), \bm{\upTheta}(t), s \right)
= \max_{c \in \mathcal{C}(t)} \left\{ \widehat{\tau}_{0}(\bm{\uptheta}_{c}(t)) \right\} \nonumber \\
& \qquad + \littlesum_{c \in \mathcal{C}(t)} \widehat{\tau}_{1}(\bm{\uptheta}_{c}(t))
+ \frac{M}{s} \cdot \littlesum_{c \in \mathcal{C}(t)} \widehat{\tau}_{2}(\bm{\uptheta}_{c}(t))
+ R_{\text{p99.9}},
\label{eq:tail-latency-model}
\end{align}
where $\mathcal{C}(t)$ denotes the set of heterogeneous cells to be served in slot $t$,
$\bm{\uptheta}_{c}(t)$ denotes the configuration for cell $c \in \mathcal{C}(t)$, and
$R_{\text{p99.9}}$ denotes the safety reserve introduced to account for the profiled median-to-p99.9 gap as well as residual model error.
One can interpret {\eqref{eq:tail-latency-model}} as follows:
When a slot contains multiple heterogeneous cells, their setup overheads can overlap; the model therefore conservatively takes the largest configuration-specific setup coefficient.
In contrast, every scheduled cell contributes processing work, so the model sums the SM-insensitive and SM-parallel coefficients across cells.
Note that the model depends only on the number of allocated SMs, $s \leq M$, not on which physical SMs are selected.
As a result, only the SM-parallel term varies with $s$.
Requiring the predicted p99.9 processing latency to satisfy the deadline $L_0$ gives
\begin{align}
& \widehat{L}_{\text{p99.9}} \left( \mathcal{C}(t), \bm{\upTheta}(t), s \right) < L_0 \Leftrightarrow \nonumber \\
& \quad s > \frac{M \cdot \littlesum_{c} \widehat{\tau}_{2}(\bm{\uptheta}_{c}(t))}{L_0 - \left( \max_{c} \left\{ \widehat{\tau}_{0}(\bm{\uptheta}_{c}(t)) \right\} + \littlesum_{c} \widehat{\tau}_{1}(\bm{\uptheta}_{c}(t)) + R_{\text{p99.9}} \right)},
\label{eq:required-vran-sms}
\end{align}
where $L_0 = 1.5\,\text{msec}$ is the 3-slot uplink processing deadline for numerology~1.
The estimated SM count is then rounded up to the smallest supported, pre-built SM allocation.

%% file: sections/05_implementation.tex
\section{Experimental Setup}
\label{sec:impl}

\begin{table}[!t]
    \centering
    \caption{Hardware platforms used in our experiments.}
    \label{tab:eval-platforms}
    \vspace{-3.0mm}
    \footnotesize
    \setlength{\tabcolsep}{4.5pt}
    \renewcommand{\arraystretch}{1.05}
    \begin{tabular}{l c c c}
        \toprule
        \textbf{Setup} &
        \textbf{CPU} &
        \textbf{GPU} &
        \textbf{Memory (bandwidth)} \\
        \midrule
        \myCodeShort{srv1} &
        EPYC 9355 &
        H200 &
        141\,GB HBM3e (4.8\,TB/s) \\

        \myCodeShort{srv2} &
        Xeon Gold 6348 &
        A100 &
        40\,GB HBM2 (1.6\,TB/s) \\

        DGX Spark &
        \multicolumn{2}{c}{GB10 Superchip} &
        128\,GB LPDDR5X (273\,GB/s) \\

        GH200 &
        \multicolumn{2}{c}{Grace Hopper Superchip} &
        96\,GB HBM3 (4.0\,TB/s) \\
        \bottomrule
    \end{tabular}
    \vspace{-2.0mm}
\end{table}

\myparatight{Implementation and hardware platforms.}
We build {\sys} on NVIDIA Aerial CUDA-accelerated RAN~\cite{nvidia_aerial} (release 25-3); our implementation adds 8.4K lines of code across {\sys}'s three modules.
Table~\ref{tab:eval-platforms} summarizes the hardware platforms used in our experimental evaluation.
Most experiments are conducted on server~\#1 (\myCodeShort{srv1}) with an H200 (Hopper) GPU, and our cross-GPU study considers server~\#2 (\myCodeShort{srv2}) with an A100 (Ampere) GPU, a DGX Spark with a GB10 (Grace-Blackwell) Superchip, and a GH200 (Grace Hopper) Superchip.
For a GPU with $M$ SMs, we use \emph{an $s$:$(M-s)$ SM split} to denote an allocation of $s$ SMs to vRAN and the remaining $(M-s)$ SMs to the co-tenant.

\myparatight{vRAN test vectors and vRAN workloads.}
We consider a total of 28 cell configurations, or \emph{test vectors}, summarized in Appendix~\ref{sec:appendix-tv} (Table~\ref{tab:tvintensity}), in 5G numerology 1 with 30\,kHz subcarrier spacing and 0.5\,ms slot duration.
To isolate compute and memory interference, we focus on a \emph{compute-heavy} test vector (\myCodeShort{tv-21}) with the highest per-slot processing cost and a \emph{memory-heavy} test vector (\myCodeShort{tv-28}) with the highest per-cell DRAM traffic.
Both configurations use a {100}~\thinspace{MHz} channel with 273 physical resource blocks (PRBs) and four antennas, but differ in the MCS.
Because \myCodeShort{tv-21} and \myCodeShort{tv-28} place the greatest observed pressure on the target GPU resources, they provide conservative stress tests for {\sys}.

\myparatight{Latency profiling and modeling.}
We derive the latency model in \S\ref{ssec:design-latency-model} offline using homogeneous vRAN \emph{uplink} workloads, in which all cells scheduled in a slot have the same configuration.
For each of the 28 test vectors, we sweep six cell counts and six vRAN SM budgets, yielding 1,008 configurations.
We run each workload for 3,000 slots and use the resulting end-to-end latency measurements to fit the latency model coefficients and calibrate the p99.9 safety reserve, as described in \S\ref{ssec:design-latency-model}.
At runtime, {\sys} composes the fitted coefficients to predict the tail (p99.9) latency of vRAN workloads with heterogeneous cells and selects the minimum SM allocation that satisfies the deadline. 
Appendix~\ref{sec:appendix-tv} provides further details on the profiling and calibration procedure.

\myparatight{Co-tenant workloads.}
We organize the co-tenant workloads into three classes that provide progressively greater realism.
First, the \emph{single-axis stress-test kernels} isolate the two principal sources of GPU contention: FMA (fused multiply-add) Burn generates controllable compute pressure through a tunable kernel duration, whereas HBM Flooder streams data past the caches to generate sustained memory bandwidth pressure.
Their controlled and independently tunable resource demands allow us to evaluate the effectiveness of individual modules of {\sys}.
Second, the \emph{production inference kernels}, including CUTLASS GEMM~\cite{nvidia_cutlass}, FlashAttention-3~\cite{shah2024flashattention3}, Triton fused MoE~\cite{tillet2019triton}, and FlashInfer~\cite{ye2025flashinfer}, represent common building blocks of modern AI inference, covering dense matrix multiplication, self-attention, expert feed-forward computation, and paged attention.
They expose realistic combinations of compute and memory demand while remaining sufficiently focused to explain kernel-level interference.
Finally, the \emph{full-stack serving workloads} use vLLM~\cite{kwon2023vllm} to decode tokens with Llama-3.3-70B~\cite{llama3} and Mistral-Small-24B.
These workloads combine compute- and bandwidth-intensive kernels with the scheduling, memory management, and runtime behavior of an end-to-end LLM serving stack.

\myparatight{Baseline configurations.}
We evaluate {\sys} using both controlled mechanism-level microbenchmarks and comparisons with prior GPU sharing systems.
We focus on the following baseline configurations:
\begin{itemize}[leftmargin=*, topsep=2pt, itemsep=1pt]
    \item 
    \textbf{vRAN only}, as a contention-free reference, runs the vRAN workload in its assigned GC partition \emph{without a co-tenant};
    \item
    \textbf{GC only} assigns the vRAN and co-tenant disjoint SM partitions via GC but disables the \emph{Cooperative HBM Bandwidth Guard}, thereby leaving HBM bandwidth shared.
    \item
    \textbf{{\sys}} uses the same SM split enabled by GC and additionally enables the cooperative HBM bandwidth guard.
\end{itemize}
To microbenchmark {\sys}'s protection mechanisms, we include two further ablations:
\textbf{Unprotected} co-schedules both workloads on the same SMs without either protection mechanism, whereas \textbf{HBM guard only} enables the cooperative HBM bandwidth guard without GC-based SM isolation.

\myparatight{Performance metrics.}
For the vRAN workload, we measure per-slot GPU execution time from device-side traces collected with NVIDIA Nsight Systems~\cite{nvidia_nsight_systems}.
For each experiment, we collect per-slot \emph{processing latency} measurements over 50,000 slots, from which the processing latency distribution, including the median (p50) and tail (p99.9), is derived.
We also report the \emph{deadline miss rate}, defined as the fraction of slots violating the 1.5\,ms processing deadline for the uplink (3-slot in numerology~1).
For production inference kernels, we report throughput as the number of completed kernel executions per second (kernels/s).
For full-stack serving applications using vLLM, we report the \emph{serving throughput} as the number of output tokens produced per second (tok/s).

%% file: sections/06_evaluation.tex
\section{Evaluation Results}
\label{sec:eval}

\begin{figure}[!t]
\centering
\includegraphics[width=0.95\columnwidth]{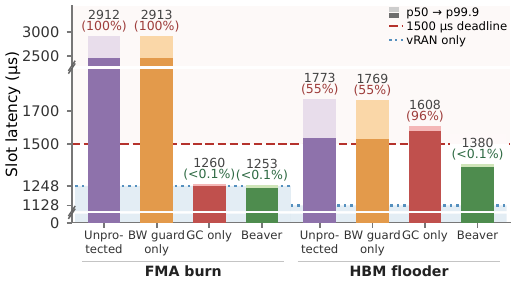}
\vspace{-3.0mm}
\caption{{\sys}'s vRAN latency protection mechanisms against single-axis stressor co-tenants: 14 compute-heavy cells with an FMA burner (104:28 SM split), and 14 memory-heavy cells with an HBM flooder (72:60 SM split).}
\label{fig:eval-single-axis-stressors}
\vspace{-3.0mm}
\end{figure}

\begin{figure*}[!t]
\centering
\includegraphics[width=0.90\textwidth]{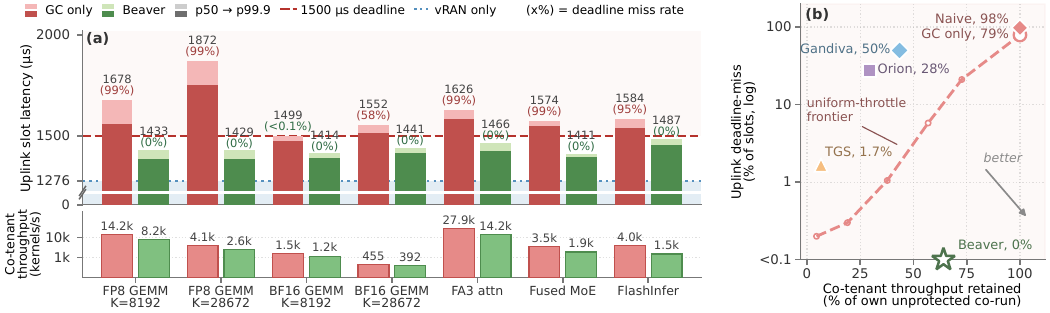}
\vspace{-3.0mm}
\caption{(a) vRAN processing latency (p50 and p99.9) with production inference kernels under \emph{GC only} and {\sys}, using 16 memory-heavy cells under a 72:60 SM split.
(b) Deadline miss rate vs. retained co-tenant (\myCodeShort{FP8} GEMM) throughput for 23 memory-heavy cells under a 104:28 SM split, comparing {\sys} with prior GPU sharing systems and baselines.}
\label{fig:eval-kernels-and-comparison}
\vspace{-2.0mm}
\end{figure*}

\subsection{Microbenchmarks}
\label{ssec:eval-microbenchmarks}

\myparatight{Sharing with single-axis stressors.}
We first isolate the effectiveness of {\sys}'s individual mechanisms using the two single-axis stressors. 
We pair 14 compute-intensive cells (\myCodeShort{tv-21}) with the FMA burner under a 104:28 SM split, and 14 memory-intensive cells (\myCodeShort{tv-28}) with the HBM flooder under a 72:60 SM split.
Fig.~\ref{fig:eval-single-axis-stressors} reports the p50 and p99.9 vRAN processing latency, with deadline miss rates shown in parentheses.
With the FMA burner, \emph{Unprotected} produces a p99.9 latency of 2.912\,ms and a 100\% miss rate, regardless of whether the HBM bandwidth guard is enabled.
\emph{GC only} reduces p99.9 latency to 1.260\,ms, nearly matching the 1.248\,ms latency of \emph{vRAN only} without the co-tenant, showing that SM isolation is sufficient to eliminate compute contention.

With the HBM flooder, however, neither mechanism alone satisfies the deadline: \emph{Unprotected} and \emph{HBM guard only} yield a p99.9 latency of 1.773\,ms and 1.769\,ms, respectively, while \emph{GC only} yields 1.608\,ms.
By combining SM isolation with cooperative HBM regulation, {\sys} reduces p99.9 latency to 1.380\,ms and the deadline miss rate to below 0.1\%.
These results show that SM partitioning isolates compute contention, whereas protecting vRAN from memory-intensive co-tenants requires coordinated SM isolation and HBM bandwidth control.
The remaining experiments compare {\sys} with only the \emph{GC only} baseline; the other baselines (\emph{Unprotected} and \emph{HBM guard only}) already fail to protect the vRAN deadline due to non-isolated compute resources.

\myparatight{Sharing with production inference kernels.}
We next evaluate {\sys} against production inference kernels that place substantial pressure on HBM: CUTLASS GEMMs, Flash\-Attention-3, fused MoE, and Flash\-Infer.
For the GEMMs, we evaluate \myCodeShort{FP8} and \myCodeShort{BF16} at two dimensions used by Llama-3.3-70B~\cite{llama3}.
$K=8{,}192$ is the model's hidden width and is used by its attention and MLP gate/up projections, whereas $K=28{,}672$ is the intermediate width consumed by the MLP down projection.
The vRAN workload comprises 16 memory-heavy cells (\myCodeShort{tv-28}) and runs under a 72:60 SM split, determined by {\sys}'s workload-aware SM sizer.
Fig.~\ref{fig:eval-kernels-and-comparison}(a) shows the vRAN processing latency and co-tenant throughput under \emph{GC only} and {\sys}.
Without the co-tenant, the vRAN's p99.9 processing latency is 1.276\,ms, as indicated by the blue dotted line (\emph{vRAN only}).
Under \emph{GC only}, six of the seven co-tenant kernels push the vRAN's median latency beyond the 1.5\,ms deadline, producing deadline miss rates of 58\%--99\%. The higher \myCodeShort{FP8} inflation is consistent with its deeper staging pipeline, which keeps more HBM transfers in flight than \myCodeShort{BF16}.
{\sys} keeps p99.9 latency within the deadline and eliminates all observed deadline misses, while the co-tenant kernels retain 38\%--86\% of their unthrottled throughput.

\myparatight{Comparison with prior GPU sharing systems.}
We compare {\sys} with Orion~\cite{strati2024orion}, TGS~\cite{wu2023tgs}, and Gandiva~\cite{xiao2018gandiva}, prior GPU sharing systems designed for \emph{general ML workloads} rather than latency-critical vRAN workloads (see Table~\ref{tab:comparison-related-work}).
We consider 23 memory-heavy cells with an \myCodeShort{FP8} GEMM co-tenant, with a 104:28 SM split selected by {\sys}'s SM sizer.
Fig.~\ref{fig:eval-kernels-and-comparison}(b) plots each system's deadline miss rate against the retained co-tenant throughput.
The dashed curve shows a uniform-throttling frontier obtained by repeatedly running the co-tenant for a fixed interval and then idling it for a configurable interval.
A system improves on this frontier only if it retains more throughput at the same deadline miss rate, or achieves a lower miss rate at the same throughput.
Sweeping the idle interval produces different throughput and deadline miss rate trade-offs.

Prior systems remain constrained by this tradeoff, reflecting that they were not designed to protect workloads with strict per-slot latency requirements.
\emph{Unprotected} sharing and \emph{GC only} retain nearly all co-tenant throughput but miss 98\% and 79\% of the slot deadlines, respectively.
Gandiva and Orion reduce the miss rate to 50\% and 28\% while retaining less than half of the co-tenant's throughput, whereas TGS reaches a 1.7\% miss rate only by reducing retained throughput below 10\%.
Even the safest uniform-throttling point remains above 0.1\% miss rate.
In contrast, {\sys} is the only system that occupies the favorable bottom-right region beyond this frontier, incurs no observed deadline misses, and retains 64.1\% of the co-tenant's unthrottled throughput.

\begin{table}[!t]
\centering
\caption{MPS-based GPU sharing systems (e.g., YinYangRAN~\cite{yinyangran}) \emph{fails} to protect the vRAN latency.}
\label{tab:comparison-mps}
\vspace{-3.5mm}
\footnotesize
\setlength{\tabcolsep}{4.0pt}
\renewcommand{\arraystretch}{1.05}
\begin{tabular}{l c c c c}
\toprule
\textbf{Co-tenant} & \textbf{\specialcell{HBM \\ Demand}} & \textbf{\specialcell{Latency \\(p50)}} & \textbf{\specialcell{Latency \\(p99.9)}} & \textbf{\specialcell{Deadline \\ Miss}} \\
\midrule
RNN forecaster~\cite{yinyangran} & 0.6\,GB/s & 1308 & 1340 (\cmark) & 0\% \\
RNN, SM-saturating & 873\,GB/s & 1247 & 1354 (\cmark) & 0.02\% \\
HBM flooder & 3{,}888\,GB/s & 1634 & 1961 (\xmark) & 72.6\% \\
FP8 GEMM & 3{,}648\,GB/s & 1608 & 1730 (\xmark) & 99.98\% \\
\bottomrule
\end{tabular}
\vspace{-2.0mm}
\end{table}

\myparatight{Comparison with YinYangRAN.}
YinYangRAN~\cite{yinyangran} is the closest prior system for sharing a GPU between vRAN and co-tenant workloads.
We reproduce its MPS-based active-thread partitioning and vary only the co-tenant workload.
As shown in Table~\ref{tab:comparison-mps}, the (lightweight) RNN forecaster used in YinYangRAN consumes only 0.6\,GB/s of HBM bandwidth and causes no vRAN deadline misses. 
Even after scaling the RNN to saturate its assigned SM capacity, the vRAN maintains a p99.9 latency of 1.354\,ms, with a 0.02\% deadline miss rate.
However, the same MPS partition fails under bandwidth-intensive co-tenants: an HBM flooder and an LLM decode-like FP8 GEMM increase the miss rate to 72.6\% and 99.98\%, respectively.
The fundamental limitation is that MPS does not isolate or regulate HBM bandwidth.
Moreover, MPS introduces substantially higher runtime overhead than the GC-based SM partitioning used by {\sys} (see \S\ref{ssec:eval-abalation}).

%% file: sections/06b_dynamic_partition_eval.tex
\subsection{End-to-End Evaluations}
\label{ssec:eval-end-to-end}

\begin{figure}[!t]
\centering
\includegraphics[width=0.90\columnwidth]{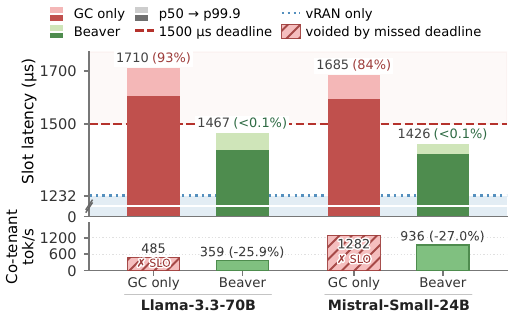}
\vspace{-3.5mm}
\caption{Uplink latency and co-tenant serving rate under a full vLLM co-tenant. Parentheses report deadline miss rates.}
\label{fig:eval-vllm}
\vspace{-3.0mm}
\end{figure}

\myparatight{Sharing with full-stack serving applications.}
For end-to-end evaluations, we first consider {\sys} with LLM inference co-tenants.
In both experiments, the vRAN workload comprises 13 memory-heavy cells (\myCodeShort{tv-28}) and runs under a 56:76 SM split.
Fig.~\ref{fig:eval-vllm} reports the vRAN processing latency and ML serving throughput under \emph{GC-only} and {\sys}.
Under \emph{GC-only}, contention from either ML serving workload pushes the vRAN's p50 processing latency beyond the 1.5\,ms deadline, resulting in deadline miss rates of 93\% with Llama-3.3-70B and 84\% with Mistral-Small-24B.
In contrast, {\sys} keeps p99.9 latency at 1.467 and 1.426\,ms, respectively, with miss rates below 0.1\%, while retaining 74\% and 73\% of the co-tenant's unthrottled serving throughput.

\begin{figure}[!t]
\centering
\includegraphics[width=0.90\linewidth]{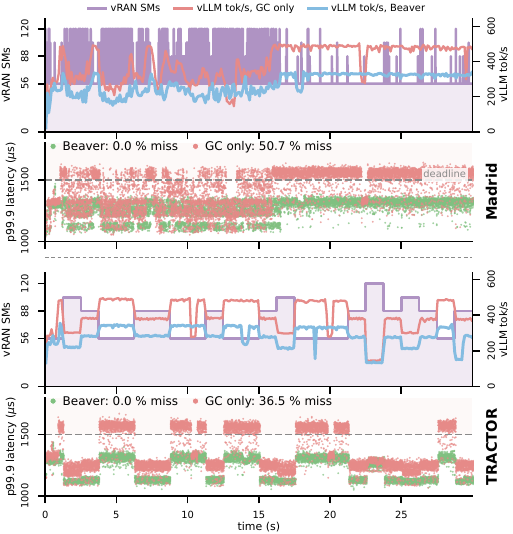}
\vspace{-3.0mm}
\caption{{\sys} adapts the vRAN SM allocation under the time-varying \emph{Madrid} (top) and \emph{O-RAN TRACTOR} (bottom) traces while co-serving Llama3.3-70B using vLLM.}
\label{fig:eval-tracereplay}
\vspace{-3.0mm}
\end{figure}

\myparatight{{\sys} under dynamic vRAN workloads.}
Next, we evaluate whether {\sys} can preserve the vRAN deadline under time-varying workloads and reclaim unused GPU capacity for a live vLLM co-tenant serving Llama3.3-70B.
We replay two public traces that capture complementary sources of variation.
The \emph{Madrid} dataset records millisecond-granularity traffic from a commercial base station~\cite{perez2023madrid}. 
We select the single base station trace and map its traffic intensity to vRAN demand, and replicate the resulting workload across 16 cells.
The \emph{O-RAN TRACTOR} dataset records the time-varying MCS observed by four user devices~\cite{groen2024tractor}.
Assuming full PRB utilization, we replicate each device's MCS trace across four cells, yielding a 16-cell workload in which temporal variation is driven by the measured MCS changes.

Fig.~\ref{fig:eval-tracereplay}(a) shows the rapidly varying vRAN workload based on the \emph{Madrid} trace, where the SM allocation determined by {\sys} adapts accordingly throughout the replay.
The top panel shows the vRAN's SM allocation and the resulting vLLM throughput, while the bottom panel compares the vRAN's p99.9 latency under \emph{GC only} and {\sys}.
Fig.~\ref{fig:eval-tracereplay}(b) reveals a steadier variation under the \emph{TRACTOR} workload, making the resource tradeoff between the tenants more clear: allocating more SMs to the vRAN in response to increasing vRAN workloads reduces vLLM throughput, whereas a decrease in vRAN workload releases SMs to the co-tenant and allows its throughput to recover.
Across both 30-second trace replays, {\sys} incurs no deadline misses, whereas \emph{GC only} misses the deadline for 37\%--51\% of the slots.
These results show that {\sys} adapts at \emph{slot granularity} to time-varying vRAN workloads and allows the vLLM co-tenant to exploit idle capacity without violating vRAN's deadline.

%% file: sections/06c_abalation.tex
\subsection{Ablation Studies}
\label{ssec:eval-abalation}

\begin{figure}[!t]
\centering
\includegraphics[width=0.95\columnwidth]{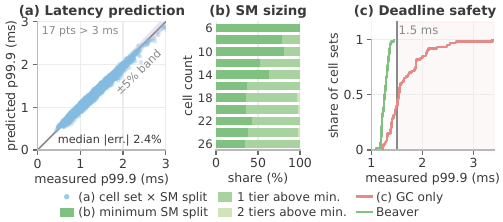}
\vspace{-3.5mm}
\caption{{\sys}'s latency model and workload-aware SM sizer: (a) Predicted vs. measured p99.9 latency, (b) Selected SM allocation vs. measured minimum, and (c) p99.9 latency under an FP8 GEMM co-tenant.}
\label{fig:eval-sm-sizer}
\vspace{-3.0mm}
\end{figure}

\myparatight{Processing latency model and SM sizer.}
The coefficients of {\sys}'s latency model in {\eqref{eq:tail-latency-model}}, $\{ \widehat{\tau}_0(\bm{\uptheta}), \widehat{\tau}_1(\bm{\uptheta}), \widehat{\tau}_2(\bm{\uptheta}) \},$ are fitted using only vRAN workloads with \emph{homogeneous} cells (see \S\ref{sec:impl}).
To evaluate how well the model generalizes compositionally, we construct vRAN workloads with 4,000 heterogeneous configurations by randomly sampling the cell count, the test vector of each cell, and the SM allocation.
Fig.~\ref{fig:eval-sm-sizer}(a) compares the predicted and measured p99.9 processing latency.
{\sys} achieves a median relative prediction error of only 2.4\%, with a 95th percentile error of 9.2\%.
Fig.~\ref{fig:eval-sm-sizer}(b) compares the SM allocation selected by {\sys} with the ground truth minimum, obtained by exhaustively searching through the supported SM allocations.
It can be seen that {\sys}'s SM sizer never underprovisions: across all cases, it selects either the exact minimum or budgets one or two allocation tiers above it (eight SMs per tier).
Finally, Fig.~\ref{fig:eval-sm-sizer}(c) evaluates deadline safety across 100 randomly sampled heterogeneous vRAN workloads at the selected SM allocation with an \myCodeShort{FP8} GEMM co-tenant. 
With \emph{GC only}, 62 of the 100 vRAN workloads exceed the 1.5\,ms deadline while {\sys} keeps all workloads within the deadline.
These results show that {\sys}'s latency model fitted only on homogeneous workloads can accurately size unseen heterogeneous workloads while preserving tail latency safety under co-tenancy.

\myparatight{Slot-granular SM repartitioning.}
We also measure the latency of changing the SM allocation across 500 repeated trials.
Using a pool of pre-created GC pairs, {\sys} switches to a new SM partition in only 1.79\,{\us}, less than 0.5\% of a 500\,{\us{}} slot. 
In contrast, reconfiguring the SM allocation takes on average 564\,ms with MPS~\cite{nvidia_mps} and 1.06\,s with MIG~\cite{nvidia_mig}, exceeding the slot duration by more than 1,000$\times$ (Fig.~\ref{fig:switch-latency}).

\begin{figure}[!t]
\centering
\includegraphics[width=0.90\linewidth]{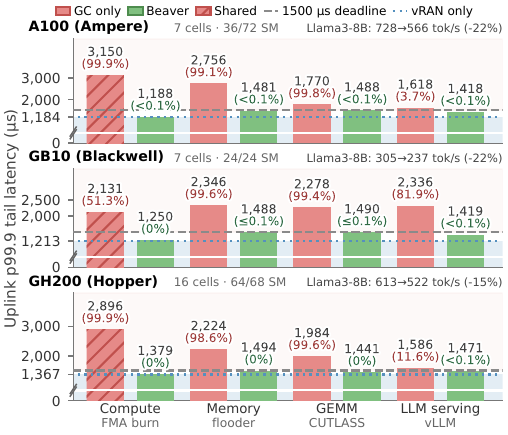}
\vspace{-4.0mm}
\caption{{\sys} on A100, GB10 and GH200: p99.9 processing latency and deadline miss rates across four co-tenants.}
\label{fig:eval-cross-gpu}
\vspace{-3.0mm}
\end{figure}

\myparatight{{\sys} across GPU architectures.}
We run {\sys} on an A100 GPU (\myCodeShort{srv2}), a GB10 Superchip (DGX Spark) and a Grace Hopper Superchip (GH200).
The vRAN workload comprises seven cells on A100 and GB10 and 16 cells on GH200. We allocate 36 of the A100’s 108 SMs, 24 of the GB10’s 48 SMs, and 64 of the GH200’s 132 SMs to the vRAN, leaving the remaining SMs to the co-tenant. 
Fig.~\ref{fig:eval-cross-gpu} shows the vRAN latency when sharing the GPU with the two single-axis stressors, a GEMM kernel, and a vLLM serving Llama-3-8B-Instruct~\cite{llama3}.
Under \emph{GC-only}, every co-tenant workload pushes the p99.9 processing latency beyond the deadline.
{\sys} keeps the vRAN's p99.9 latency within the deadline and reduces the deadline miss rates to below 0.1\%, while retaining 78\%--85\% of the unthrottled vLLM serving throughput on all platforms.

\begin{figure}[!t]
\centering
\includegraphics[width=0.90\linewidth]{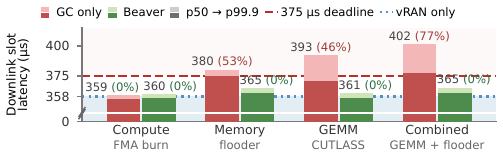}
\vspace{-4.0mm}
\caption{{\sys} for vRAN downlink workload with different compute- and memory-intensive co-tenants.}
\label{fig:eval-abalation-downlink}
\vspace{-3.0mm}
\end{figure}

\myparatight{{\sys} for downlink processing.}
We also evaluate whether {\sys}'s protection mechanisms carry over for downlink processing with a tighter deadline of 375\,{\us}.
Fig.~\ref{fig:eval-abalation-downlink} shows that for vRAN downlink workload, \emph{GC only} suffices against the compute-intensive FMA burner, keeping p99.9 latency within deadline.
However, the HBM flooder, CUTLASS GEMM, and their combination increase p99.9 latency above the processing deadline, producing deadline miss rates of 53\%, 46\%, and 77\%, respectively.
With HBM guard enabled, {\sys} keeps p99.9 latency below the deadline in every case and eliminates all observed deadline misses, confirming that it generalizes effectively to downlink processing.

%% file: sections/07_realted_work.tex
\section{Related Work}
\label{sec:rw}

\myparatight{Baseband processing and virtualization.}
Sora~\cite{tan2011sora} demonstrated high-performance wireless processing on general-purpose CPUs, and BigStation~\cite{yang2013bigstation}, Agora~\cite{ding2020agora, qi2023programmable}, and Hydra~\cite{gong2023scalable} scaled software MIMO processing within and across servers. 
Savannah~\cite{qi2024savannah, qi2024savannah2} and Nexus~\cite{qi2025nexus} explore heterogeneous CPU--eASIC processing, while Intel FlexRAN, srsRAN, and OpenAirInterface demonstrate vRAN on commodity infrastructure~\cite{intel_flexran,srsran,OAI}.
These systems make software L1 feasible and scalable, but do not protect a latency-critical GPU L1 pipeline from concurrent ML workloads.

\myparatight{GPU-accelerated L1.}
Prior work accelerates individual L1 operations, particularly LDPC decoding, on GPUs~\cite{tarver2021gpu,ling2021fast, zheng2026real}. 
DecodeX~\cite{qi2026decodex} benchmarks LDPC decoding across CPU, GPU, and ASIC platforms, exposing the
effects of parallelism, workload granularity, data movement, and offload.
NVIDIA Aerial instead implements complete, production-grade PDSCH and PUSCH pipelines with CUDA
kernels~\cite{nvidia_aerial}. 

\myparatight{GPU sharing systems (see Table~\ref{tab:comparison-related-work}).}
Gandiva, AntMan, and HiveD schedule or resize training jobs~\cite{xiao2018gandiva,antman2020,hived2020};
Salus, Orion, and TGS share GPUs at iteration, operator, or kernel granularity~\cite{yu2020salus,strati2024orion,wu2023tgs}.
REEF provides {\us}-scale preemption, but requires a custom AMD ROCm driver and is incompatible with our CUDA pipeline~\cite{han2022reef}.
MIG isolates SMs and memory slices, whereas MPS and CUDA green contexts isolate compute but leave HBM shared~\cite{nvidia_mig,nvidia_mps,nvidia_greenctx}.
For vRAN systems, YinYangRAN shares compute through MPS, while CAORA and RAN-LLM use MIG to colocate vRAN with AI or LLM workloads~\cite{yinyangran,caora,llm_ran_cloud} (Fig.~\ref{fig:motivation-failures}(d)).

%% file: sections/08_conclusion.tex
\section{Conclusion}
\label{sec:conclusion}

We present {\sys}, a GPU sharing system that enables elastic co-tenancy between ML and latency-critical vRAN workloads.
{\sys} combines a workload-aware SM sizer for heterogeneous multi-cell workloads, slot-granular SM repartitioning across pre-created GC pairs, and a cooperative HBM bandwidth guard injected into compiled co-tenant kernels.
Evaluated using NVIDIA Aerial, {\sys} keeps the vRAN's p99.9 processing latency within its deadline while co-located with full-stack vLLM serving, and extends effectively to the downlink pipeline and other GPUs.
Future directions include coordinating multiple ML co-tenants, extending resource management across multi-GPU deployments, and exploring hardware support for finer-grained HBM bandwidth control.

%% file: sections/09_appendix.tex
\section{Aerial Uplink Test Vector Selection}
\label{sec:appendix-tv}

\myparatight{What is a test vector?}
An Aerial test vector is a reproducible input for one execution of the radio
pipeline. Each uplink vector represents one 0.5-ms slot and stores the received
frequency-domain samples, the radio configuration---including its resource
blocks, symbols, modulation and coding scheme (MCS), code rate, MIMO layers,
and reference-signal configuration---and reference outputs for checking the
decoded result. The benchmark replays this slot and replicates it across cells.
A test vector therefore defines the work for one cell in one slot; it is not a
multi-slot traffic trace.

\myparatight{Which vectors are available?}
Aerial provides uplink vectors with different bandwidth allocations, antenna
configurations, user and layer counts, symbol allocations, and MCS values. We
choose the four-receive-antenna, four-layer family because it is the only
supplied family that covers every MCS value, 0--27, while also using the maximum
carrier bandwidth and maximum single-user layer count. This family contains 35
vectors on a 100-MHz carrier with 273 resource blocks and one user. Each vector
captures one complete 0.5-ms uplink slot. With 30-kHz subcarrier spacing and a
normal cyclic prefix, one slot has 14 OFDM symbols, and these vectors assign all
14 to PUSCH. The 16-symbol configuration instead counts uplink symbols across a DDDSU pattern: 14 from the full uplink slot and two from the special slot. Because our benchmark replays only the full uplink slot, each test vector contains 14 symbols.

Table~\ref{tab:tvintensity} lists the family in MCS order, one row per MCS. The first group of columns
describes each radio configuration. The remaining columns report its measured
compute and HBM demand; darker shading means greater demand within that column.
We profile every vector using the same 20-cell run and a fixed 72-SM uplink
partition. The two highlighted rows, vectors~21 and~28, are the
compute-heavy and memory-heavy vectors used in our evaluation.

\begin{table*}[t]
\centering
\footnotesize
\setlength{\tabcolsep}{4pt}
\caption{Radio configuration and measured GPU demand of the \uplink{}
test-vector suite, one row per MCS (3GPP MCS Table~2~\cite{tgpp_38214}).
Darker shading marks higher demand within a measured column; the two
highlighted rows are the vectors the evaluation uses.}
\label{tab:tvintensity}
\vspace{-3.0mm}
\begin{tabular*}{\dimexpr\textwidth-2\tabcolsep\relax}{@{\extracolsep{\fill}}l c c r r r r r r r r@{}}
\toprule
 & & & & & & \multicolumn{5}{c}{Measured GPU demand} \\
\cmidrule(lr){7-11}
\textbf{Test Vector ID (\myCodeShort{tv-ID})} & \textbf{MCS} & \textbf{Mod.} & \shortstack{Code \\ rate} & \shortstack{Code \\ Blocks} & \shortstack{Rate (Gbps)} & \shortstack{Compute \\($\mu$s/cell)} & \shortstack{SM\\(\% peak)} & \shortstack{HBM traffic\\(MB/cell)} & \shortstack{Sustained HBM\\(GB/s)} & \shortstack{Peak HBM\\(GB/s)} \\
\midrule
1 & 0 & \textbf{QPSK} & 0.117 & 11 & 0.080 & \cellcolor[HTML]{FFCDC2} 62.4 & \cellcolor[HTML]{FFEBE1} 46 & \cellcolor[HTML]{EBF0FF} 8.0 & \cellcolor[HTML]{E1EAFF} 128 & \cellcolor[HTML]{EBF0FF} 336 \\
2 & 1 &  & 0.188 & 17 & 0.129 & \cellcolor[HTML]{FFC4B8} 67.1 & \cellcolor[HTML]{FFC6BA} 52 & \cellcolor[HTML]{E9EFFF} 8.1 & \cellcolor[HTML]{EBF0FF} 121 & \cellcolor[HTML]{DAE6FF} 397 \\
3 & 2 &  & 0.301 & 13 & 0.205 & \cellcolor[HTML]{FFE5DB} 51.0 & \cellcolor[HTML]{FFC5BA} 52 & \cellcolor[HTML]{E9EFFF} 8.1 & \cellcolor[HTML]{BBD4FF} 159 & \cellcolor[HTML]{E6EDFF} 353 \\
4 & 3 &  & 0.438 & 18 & 0.295 & \cellcolor[HTML]{FFE8DE} 49.3 & \cellcolor[HTML]{FFC2B7} 52 & \cellcolor[HTML]{E7EEFF} 8.3 & \cellcolor[HTML]{B0CEFF} 167 & \cellcolor[HTML]{81B2FF} 725 \\
5 & 4 &  & 0.588 & 24 & 0.402 & \cellcolor[HTML]{FFEBE1} 48.3 & \cellcolor[HTML]{FFBCB0} 53 & \cellcolor[HTML]{E4ECFF} 8.5 & \cellcolor[HTML]{A6C7FF} 176 & \cellcolor[HTML]{73AAFF} 779 \\
\addlinespace[2pt]
\specialrule{0.7pt}{1.5pt}{1.5pt}
6 & 5 & \textbf{16-QAM} & 0.369 & 30 & 0.500 & \cellcolor[HTML]{FFADA1} 77.8 & \cellcolor[HTML]{FFB0A4} 55 & \cellcolor[HTML]{D7E4FF} 9.5 & \cellcolor[HTML]{E9EFFF} 122 & \cellcolor[HTML]{C0D7FF} 493 \\
7 & 6 &  & 0.424 & 35 & 0.574 & \cellcolor[HTML]{FFBBB0} 71.1 & \cellcolor[HTML]{FF8073} 63 & \cellcolor[HTML]{D5E3FF} 9.7 & \cellcolor[HTML]{D8E5FF} 136 & \cellcolor[HTML]{BCD4FF} 509 \\
8 & 7 &  & 0.479 & 39 & 0.656 & \cellcolor[HTML]{FFB8AC} 72.8 & \cellcolor[HTML]{FF8F82} 61 & \cellcolor[HTML]{D4E2FF} 9.8 & \cellcolor[HTML]{DAE6FF} 134 & \cellcolor[HTML]{B7D1FF} 527 \\
9 & 8 &  & 0.540 & 44 & 0.738 & \cellcolor[HTML]{FFB4A8} 74.4 & \cellcolor[HTML]{FF9C8F} 59 & \cellcolor[HTML]{D0E0FF} 10.0 & \cellcolor[HTML]{D9E5FF} 135 & \cellcolor[HTML]{B5D1FF} 532 \\
10 & 9 &  & 0.602 & 49 & 0.819 & \cellcolor[HTML]{FFC0B4} 69.0 & \cellcolor[HTML]{FF7F71} 63 & \cellcolor[HTML]{CDDEFF} 10.2 & \cellcolor[HTML]{C8DBFF} 149 & \cellcolor[HTML]{ACCBFF} 566 \\
11 & 10 &  & 0.643 & 52 & 0.869 & \cellcolor[HTML]{FFBBAF} 71.3 & \cellcolor[HTML]{FF9386} 60 & \cellcolor[HTML]{CBDDFF} 10.4 & \cellcolor[HTML]{CBDDFF} 146 & \cellcolor[HTML]{AACAFF} 575 \\
\addlinespace[2pt]
\specialrule{0.7pt}{1.5pt}{1.5pt}
12 & 11 & \textbf{64-QAM} & 0.455 & 56 & 0.934 & \cellcolor[HTML]{FF8D7F} 93.5 & \cellcolor[HTML]{FF7566} 65 & \cellcolor[HTML]{BED6FF} 11.4 & \cellcolor[HTML]{E9EFFF} 122 & \cellcolor[HTML]{CADDFF} 456 \\
13 & 12 &  & 0.505 & 62 & 1.033 & \cellcolor[HTML]{FF897B} 95.6 & \cellcolor[HTML]{FF786A} 64 & \cellcolor[HTML]{BBD4FF} 11.7 & \cellcolor[HTML]{E9EFFF} 122 & \cellcolor[HTML]{C8DCFF} 462 \\
14 & 13 &  & 0.554 & 69 & 1.147 & \cellcolor[HTML]{FF8D80} 93.4 & \cellcolor[HTML]{FF7364} 65 & \cellcolor[HTML]{B7D1FF} 12.0 & \cellcolor[HTML]{E1EAFF} 128 & \cellcolor[HTML]{BFD6FF} 498 \\
15 & 14 &  & 0.602 & 74 & 1.246 & \cellcolor[HTML]{FF8D80} 93.4 & \cellcolor[HTML]{FF7E70} 63 & \cellcolor[HTML]{B4D0FF} 12.2 & \cellcolor[HTML]{DFE9FF} 130 & \cellcolor[HTML]{BDD5FF} 503 \\
16 & 15 &  & 0.650 & 80 & 1.344 & \cellcolor[HTML]{FF8C7F} 93.8 & \cellcolor[HTML]{FF7F71} 63 & \cellcolor[HTML]{AFCDFF} 12.6 & \cellcolor[HTML]{DAE6FF} 134 & \cellcolor[HTML]{B9D3FF} 519 \\
17 & 16 &  & 0.702 & 86 & 1.442 & \cellcolor[HTML]{FF988B} 88.3 & \cellcolor[HTML]{FF6D5E} 66 & \cellcolor[HTML]{AACAFF} 12.9 & \cellcolor[HTML]{CBDDFF} 146 & \cellcolor[HTML]{A9C9FF} 578 \\
18 & 17 &  & 0.754 & 92 & 1.541 & \cellcolor[HTML]{FF9B8E} 86.7 & \cellcolor[HTML]{FF7061} 66 & \cellcolor[HTML]{A7C8FF} 13.2 & \cellcolor[HTML]{C4D9FF} 152 & \cellcolor[HTML]{A7C8FF} 587 \\
19 & 18 &  & 0.803 & 98 & 1.639 & \cellcolor[HTML]{FFA296} 83.1 & \cellcolor[HTML]{FF7769} 65 & \cellcolor[HTML]{A3C6FF} 13.5 & \cellcolor[HTML]{B7D1FF} 162 & \cellcolor[HTML]{A3C6FF} 600 \\
20 & 19 &  & 0.853 & 104 & 1.737 & \cellcolor[HTML]{FFAB9E} 79.1 & \cellcolor[HTML]{FF7C6D} 64 & \cellcolor[HTML]{9FC3FF} 13.8 & \cellcolor[HTML]{A7C8FF} 175 & \cellcolor[HTML]{9FC4FF} 613 \\
\addlinespace[2pt]
\arrayrulecolor[HTML]{F6BE80}\specialrule{0.7pt}{1.5pt}{1.5pt}\arrayrulecolor{black}
\cellcolor[HTML]{FAD7A8}\textbf{21}~{\scriptsize(\textbf{compute-heavy})} & \textbf{20} & \textbf{256-QAM} & \textbf{0.667} & \textbf{107} & \textbf{1.803} & \cellcolor[HTML]{FF6455} \textbf{113.5} & \cellcolor[HTML]{FF6455} \textbf{68} & \cellcolor[HTML]{93BCFF} \textbf{14.7} & \cellcolor[HTML]{DFE9FF} \textbf{130} & \cellcolor[HTML]{C1D7FF} \textbf{489} \\
\arrayrulecolor[HTML]{F6BE80}\specialrule{0.7pt}{1.5pt}{1.5pt}\arrayrulecolor{black}
22 & 21 &  & 0.694 & 113 & 1.902 & \cellcolor[HTML]{FF695A} 111.0 & \cellcolor[HTML]{FF6657} 67 & \cellcolor[HTML]{8FBAFF} 15.1 & \cellcolor[HTML]{D8E5FF} 136 & \cellcolor[HTML]{B4D0FF} 538 \\
23 & 22 &  & 0.736 & 119 & 1.999 & \cellcolor[HTML]{FF7062} 107.3 & \cellcolor[HTML]{FF6859} 67 & \cellcolor[HTML]{8BB8FF} 15.4 & \cellcolor[HTML]{CFDFFF} 143 & \cellcolor[HTML]{B2CFFF} 543 \\
24 & 23 &  & 0.778 & 125 & 2.098 & \cellcolor[HTML]{FF7264} 106.4 & \cellcolor[HTML]{FF6A5C} 67 & \cellcolor[HTML]{85B4FF} 15.8 & \cellcolor[HTML]{C8DCFF} 148 & \cellcolor[HTML]{AECCFF} 558 \\
25 & 24 &  & 0.821 & 133 & 2.230 & \cellcolor[HTML]{FF7E70} 100.9 & \cellcolor[HTML]{FF6D5E} 66 & \cellcolor[HTML]{81B2FF} 16.1 & \cellcolor[HTML]{BAD3FF} 160 & \cellcolor[HTML]{ADCCFF} 562 \\
26 & 25 &  & 0.864 & 141 & 2.360 & \cellcolor[HTML]{FF8274} 98.9 & \cellcolor[HTML]{FF6A5C} 67 & \cellcolor[HTML]{7AAEFF} 16.6 & \cellcolor[HTML]{AFCDFF} 168 & \cellcolor[HTML]{A7C8FF} 587 \\
27 & 26 &  & 0.895 & 144 & 2.426 & \cellcolor[HTML]{FF8C7E} 94.1 & \cellcolor[HTML]{FF6C5D} 66 & \cellcolor[HTML]{79ADFF} 16.7 & \cellcolor[HTML]{A3C6FF} 178 & \cellcolor[HTML]{A5C7FF} 591 \\
\arrayrulecolor[HTML]{84BCE1}\specialrule{0.7pt}{1.5pt}{1.5pt}\arrayrulecolor{black}
\cellcolor[HTML]{C5DBE9}\textbf{28}~{\scriptsize(\textbf{memory-heavy})} & \textbf{27} & \textbf{} & \textbf{0.926} & \textbf{152} & \textbf{2.556} & \cellcolor[HTML]{FFAA9E} \textbf{79.3} & \cellcolor[HTML]{FF786A} \textbf{64} & \cellcolor[HTML]{73AAFF} \textbf{17.2} & \cellcolor[HTML]{73AAFF} \textbf{217} & \cellcolor[HTML]{93BCFF} \textbf{660} \\
\bottomrule
\end{tabular*}

\vspace{2pt}
{\footnotesize\setlength{\leftskip}{0pt}\setlength{\rightskip}{0pt}%
\setlength{\parfillskip}{0pt plus 1fil}\noindent
\emph{Code blocks} denotes
the LDPC code-block count per cell and \emph{Rate} denotes the decoded
transport-block throughput, both per cell. \emph{Compute} is the median
GPU execution time per slot divided by the number of cells. \emph{SM} is
Nsight Compute's SM-throughput percentage, time-averaged across the slot's
kernels by execution time. \emph{HBM traffic} is the data read from and
written to HBM by one cell in one slot, \emph{Sustained HBM} that volume
divided by the slot's GPU execution time, and \emph{Peak HBM} the
highest bandwidth reached by any kernel in the slot. All rows are
measured \uplink{}-solo at 20 cells on the same 72-SM partition. Antenna count (4), user count (1) and
transport-block count (1) are constant and omitted; every vector spans
273 PRBs.
In Aerial's own numbering, vector~21 is
\texttt{8251} and vector~28 is
\texttt{8214}.\par}
\end{table*}

\myparatight{Why these two vectors?}
We use two vectors because compute and HBM have different worst cases. Compute
demand is the median GPU execution time per slot divided by 20 cells. By this
measure, vector~21 is the suite's most compute-intensive
vector at 113.5\,\us{} per cell. Its MCS-20, 256-QAM configuration has a
relatively low code rate and creates substantial LDPC parity work.

Vector~28 represents the suite's most memory-intensive
MCS-27 configuration. It has the largest transport block and, in our profiling,
moves 17.2\,MB through HBM per cell, sustains 217\,GB/s across the slot, and
reaches 660\,GB/s in its busiest kernel.

We therefore use vector~21 for head-of-line blocking, compute contention,
and SM-partitioning experiments, where the radio's compute demand should be
largest. We use vector~28 for HBM-contention and bandwidth-protection
experiments, where the radio's memory demand should be largest. Each mechanism
is tested with the vector that places the most pressure on the resource it
controls.

\section{Aerial Downlink Test Vector Selection}
\label{sec:appendix-downlink-tv}

For the downlink evaluation, we use Aerial PDSCH test vector
\texttt{TVnr\_3942}, shortened to \myCodeShort{3942} in the main text. Unlike
the uplink suite, whose compute- and memory-heavy configurations occur at
different MCS values, the available downlink vectors have similar HBM bandwidth
demand. We therefore select the highest-MCS vector to represent a demanding
downlink configuration.

Vector~\myCodeShort{3942} represents a 0.5\,ms slot on a 100\,MHz FR1 carrier
with 273 physical resource blocks and 30\,kHz subcarrier spacing. It contains
one full-band UE, four transmit and four receive antennas, four MIMO layers, and
an MCS-27, 256-QAM grant. The evaluation replicates this vector across 52 cells
and assigns 72 of the H200's 132 SMs to the downlink and the remaining 60 SMs to
the co-tenant.

\section{SM Repartitioning Latency}
\label{sec:appendix-switch}

\begin{figure}[!t]
\centering
\includegraphics[width=0.98\columnwidth]{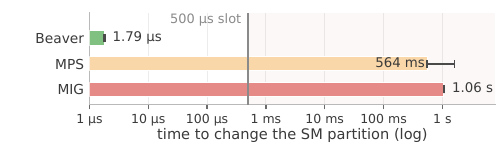}
\vspace{-3.0mm}
\caption{Time to change the SM allocation, on a log scale, against the
500\,{\us} slot.}
\label{fig:switch-latency}
\vspace{-3.0mm}
\end{figure}

\begin{figure*}[!t]
\centering
\includegraphics[width=0.98\textwidth]{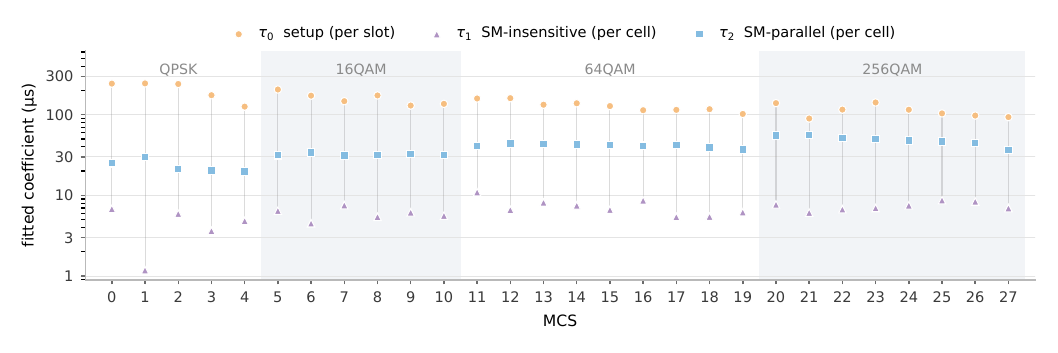}
\vspace{-6.0mm}
\caption{Fitted coefficients for 28 test vectors, one for each MCS.}
\label{fig:sizing-terms}
\vspace{-1.0mm}
\end{figure*}

We evaluate whether each partitioning mechanism can change the SM allocation within a 500\us{} radio slot.
The comparison includes {\sys}, MPS, and MIG and measures the complete procedure each mechanism requires before the new allocation can be used.

Figure~\ref{fig:switch-latency} plots the measured repartitioning latency on a logarithmic scale.
{\sys} switches to a pre-created partition in 1.79\us{}, whereas MPS and MIG require 564\,ms and 1.06\,s, respectively.
The vertical line marks the duration of one radio slot.

Only {\sys} operates at the required timescale: its switch consumes less than 0.5\% of a slot, while MPS and MIG exceed the slot duration by more than three orders of magnitude.
This confirms that pre-materializing the candidate partitions enables {\sys} to follow slot-to-slot radio demand without delaying subsequent radio processing.

\section{SM Sizing Model Calibration}
\label{sec:appendix-sizing-terms}

This appendix describes how we calibrate the sizing model in \S\ref{ssec:design-latency-model}.
Instead of profiling the combinatorial space of heterogeneous cell sets, we
measure a structured set of homogeneous workloads, in which every cell uses
the same test vector. We then compose the resulting per-vector coefficients to
predict slots containing different test vectors. No heterogeneous workload is
used for fitting; heterogeneous cell sets are used only to evaluate the model
in \S\ref{ssec:eval-abalation}.

\myparatight{Structured homogeneous sweep.}
For each of the 28 test vectors, we cross six cell counts,
$C \in \{6, 10, 14, 18, 22, 26\}$, with six vRAN SM budgets,
$s \in \{40, 56, 72, 88, 104, 120\}$. At each combination, all cells use the
same test vector, making the workload homogeneous. This produces
$6 \times 6 = 36$ combinations per test vector and
$28 \times 36 = 1{,}008$ combinations in total. We run each combination for
3,000 consecutive slots, launching one CUDA graph per slot. We discard the
first 50 warm-up slots and compute the median and p99.9 latency from the
remaining 2,950 slots. The complete sweep therefore executes 3.024 million
slots.

\myparatight{Fit each test vector.}
For each test vector, least-squares regression fits the three coefficients
$\tau_0$, $\tau_1$, and $\tau_2$ from its homogeneous measurements. We fit only
combinations whose measured median is at most 2.2\,ms. Combinations far beyond
the 1.5\,ms deadline are already infeasible and cannot affect the selected
partition; including them would reduce fitting accuracy near the deadline,
where sizing decisions matter. 

\myparatight{Predict heterogeneous slots.}
At runtime, a slot may contain cells using different test vectors. The
controller looks up the coefficients fitted for each cell's test vector, takes
the largest per-slot setup term, and sums the two per-cell terms as described
in Equation~\eqref{eq:tail-latency-model}. Thus, the structured
homogeneous combinations provide the coefficients needed to predict previously
unseen heterogeneous cell sets without profiling those combinations directly.
\S\ref{ssec:eval-abalation} evaluates this composition on held-out heterogeneous
workloads.

\myparatight{Safety reserve.}
The composed model predicts median latency, whereas partition sizing must
protect the p99.9 tail. Our calibration uses a fixed 24.6\us{}
median-to-p99.9 offset and a fitting-error allowance of 60--190\us{}, depending
on the test vector. The resulting reserve $R_{\text{p99.9}}$ is therefore
85--215\us{} for a vRAN-only slot. A heterogeneous
slot uses the largest reserve among its test vectors and adds it once to the
composed latency estimate. 

\myparatight{Fitted coefficients.}
Figure~\ref{fig:sizing-terms} shows the fit used for each of the 28 test
vectors. The per-slot setup term $\tau_0$ is highest at QPSK on average and
reaches 244\us{} there. The SM-parallel term $\tau_2$ increases from a median
of 21\us{} per cell for QPSK to 49\us{} for 256-QAM, while the SM-insensitive
term $\tau_1$ remains below 11\us{} per cell. Across all test vectors, the
median fitting error is 31\us{}, and the maximum is 76\us{}, or 5\% of the
deadline.